\documentclass[preprint]{aa}
\usepackage{graphicx}        
\usepackage{natbib}        
\usepackage{amssymb}        
\usepackage{amsmath}
\usepackage{color}           
\usepackage{url}             
\usepackage{soul}
\usepackage{MR_AA}
\usepackage{verbatim}
\usepackage{commath}
\usepackage[normalem]{ulem} 
\usepackage{soul}

\providecommand\bnabla{\boldsymbol{\nabla}}

\newcommand\bv{{\boldsymbol{v}}}
\providecommand\bnabla{\boldsymbol{\nabla}}

\def\Tau{{\rm T}}

\newcommand{\stkout}[1]{\ifmmode\text{\sout{\ensuremath{#1}}}\else\sout{#1}\fi}

\newcommand{\Tej}{T_{\rm eff}^{\rm jump}}
\newcommand{\Xc}{X_{\rm core}/X_0}

\newcommand{\Teff}{T_{\rm eff}}

\begin{document}
\titlerunning{Evolution of rotation in rapidly rotating early-type stars}
\title{Evolution of rotation in rapidly rotating early-type stars during the main sequence with 2D models}
\author{D. Gagnier\inst{1} \and M. Rieutord\inst{1} \and
C. Charbonnel\inst{2,1} \and B. Putigny\inst{1} \and F. Espinosa
Lara\inst{3}}

\institute{
IRAP, Universit\'e de Toulouse, CNRS, UPS, CNES,
14, avenue \'{E}douard Belin, F-31400 Toulouse, France
\and
Department of Astronomy, University of Geneva, Chemin des Maillettes 51,
1290, Versoix, Switzerland 
\and
Space Research Group, University of Alcal\'a, 28871 Alcal\'a de Henares, Spain
\\
\email{[Damien.Gagnier,Michel.Rieutord]@irap.omp.eu, 
Corinne.Charbonnel@unige.ch}
}

\abstract{
The understanding of the rotational evolution of early-type stars is deeply related to that of anisotropic mass and angular momentum loss. In this paper, we aim to clarify the rotational evolution of rapidly rotating early-type stars along the main sequence (MS). We have used the 2D ESTER code to compute and evolve isolated rapidly rotating early-type stellar models along the MS, with and without anisotropic mass loss. We show that stars with $Z=0.02$ and masses between $5$ and $7~M_\odot$ reach criticality during the main sequence provided their initial angular velocity is larger than 50\% of the Keplerian one. More massive stars are subject to radiation-driven winds and to an associated loss of mass and angular momentum. We find that this angular momentum extraction from the outer layers can prevent massive stars from reaching critical rotation and greatly reduce the degree of criticality at the end of the MS.
Our model includes the so-called bi-stability jump of the $\dot{M}-T_{\rm eff}$
relation of 1D-models. This discontinuity now shows up in the latitude variations of the
mass-flux surface density, endowing rotating massive stars with either a single-wind regime
(no discontinuity) or a two-wind regime (a discontinuity). In the two-winds-regime, mass loss and angular momentum loss are strongly increased at low latitudes inducing a faster slow-down of the rotation.
However, predicting the rotational fate of a massive star is difficult, mainly because of the non-linearity of the phenomena involved and their strong dependence on uncertain prescriptions. Moreover, the very existence of the bi-stability jump in mass-loss rate remains to be substantiated by observations. }

\keywords{stars: evolution -- stars: rotation -- stars: early-type -- stars: mass-loss} 

\maketitle
%

\section{Introduction}
The evolution of the rotation rate of stars is one of the open challenges of current stellar physics. The rotation rate of a star indeed depends on several un-mastered magneto-hydrodynamic mechanisms. The most important ones may be those that transport and/or extract angular momentum within the stellar interior and at the surface of the star,
and in the first place the losses due to radiation-driven winds, possibly modified by the presence of a magnetic field. Angular momentum losses depend on various phenomena but in particular on the mass loss distribution at the surface of the star. It is clear that a strong mass loss at the equator of the star is more efficient at extracting angular momentum than a strong mass loss at the pole. Moreover, the shape of a fast rotating star strongly deviates from the spherical symmetry and its spheroidal shape emphasises the anisotropy of the wind.

Until now, stellar evolution codes cope with this question using more or less sophisticated recipes. Often, the spherical symmetry of the models is coupled to a very simplified modelling of mass and angular momentum loss where the star is pealed off at a given rate \citep[e.g.][]{woosley_etal93,Ekstrom2012}. This rate may depend on some general parameters of the star (luminosity, effective temperature, etc.).

In the present work we make a step forward in the modelling of the rotational evolution of stars by using the 2D ESTER models that have been designed by \cite{ELR13} and \cite{Rieutordal2016}. These models describe the 2D steady structure of a rapidly rotating star. They can also be used to investigate  the rotational evolution of early-type stars along the main sequence, using a simple method that we implement here to compute the change in the hydrogen mass fraction in the convective core of the model, and which provides an acceptable description of the main sequence evolution. With these state-of-the-art 2D models we have now access to the latitude variations of surface quantities that matter for mass loss, namely effective gravity and effective temperature, as the star evolves. We can thus investigate in some details the consequences of the ensuing anisotropic radiation-driven winds and in particular the effects of a mass-flux jump at some effective temperature, as has been suggested by \cite{Vink2001}.

In a preliminary study, \cite{gagnier_etal19} (hereafter referred to as Paper I) have revisited the concept of critical angular velocity for stars with strong surface radiative acceleration. In this study, we first used the $\omega$-model of \cite{ELR11} to derive an analytical expression for the critical angular velocity. Briefly, the $\omega$-model assumes that radiative flux and  effective gravity are anti-parallel and that the stellar mass is centrally condensed. We show that up to 90\% of the critical angular velocity, the $\omega$-model flux does not diverge more than 10\% from the flux given by a full 2D-model. Our new expression of the critical angular velocity differs from that of \cite{MaederMeynet2000} pioneer work, and turns out to be very close to the Keplerian angular velocity for all MS evolutionary stages, at least for stellar models of mass less that $40~M_\odot$. We explain this small difference by the combined effects of gravity darkening and reduced equatorial opacity caused by the centrifugal force, and conclude that the standard Keplerian angular velocity remains a very good approximation for critical angular velocity. \cite{gagnier_etal19} also designed a local mass-flux ($\dot{m}$) and angular momentum flux prescription based on the modified CAK theory \citep{PPK} and calibrated with \cite{Vink2001} 1D models. The discontinuity included in our $\dot{m}-\Teff$ relation (the so-called bi-stability jump) makes the mass-flux of rapidly rotating star controlled by either a single-wind or a two-wind regime (respectively SWR and TWR). In the SWR, we find a maximum mass-flux at the poles, while in the TWR both mass and angular momentum losses are strongly enhanced in equatorial regions. This 2D mass loss prescription now opens the door to the present study of the main-sequence evolution of rotation in rapidly rotating early-type stars.

This paper is organised as follows. In Sect.~\ref{sec:ESTER} we present the ESTER
code and describe the nuclear time evolution framework. In Sect.~\ref{constant_angmom} we concentrate on the rotational evolution of
early-type stars assuming that mass and angular momentum losses are negligible, which is certainly valid
for intermediate-mass stars up to 
$\sim 7~M_\odot$ at metallicity close to solar. In Sect.~\ref{sectionEvolutionwithAMloss}
we consider evolution with mass loss and the associated angular momentum
loss, which is appropriate for more massive star evolution. Conclusions follow in Sect.~\ref{sec:conclu}.

\section{The ESTER code}\label{sec:ESTER}

The ESTER code self-consistently computes the steady state of an isolated rotating star. The models include the 2D axisymmetric structure and the large-scale flows (differential rotation and meridional circulation) driven by the baroclinicity of radiative regions. A short description of the ESTER models can be found in Paper I and more details in \cite{ELR13} and \cite{Rieutordal2016}.

\subsection{A simplified scheme for temporal evolution on the main sequence}

As far as time evolution is concerned, we shall concentrate in this study on the main
sequence and assume that the star remains in a quasi-steady state. We thus assume that the drivers of evolution are weak enough that all
adjustments to the steady state occur on timescales that are much
shorter than the ones imposed by the drivers, namely nuclear burning and mass loss.

Since we consider only early-type stars, nuclear reactions are located
in the convective core, which is fully mixed. Focusing on the evolution of rotation, we do not need a detailed network of nuclear reactions and simplify the evolution
of the mass fraction of hydrogen $X$ by considering its relation to the nuclear
energy production $\eps_*$. Heat being essentially
produced by the transformation of protons into $^4$He nuclei,
we assume that, locally,

\begin{equation}
\frac{\partial X}{\partial t} + \bv \cdot \bnabla X =-\frac{4 m_{\rm p} \eps_*}{Q}
\label{Xdot}
\end{equation}
where $m_{\rm p}$ is the proton mass,

\begin{equation}
Q=(4 m_{\rm p} - m_{\rm He})c^2 \simeq 4.3 \times 10^{-12} \ J
\end{equation}
is the energy released by the fusion of the four protons, and
$\eps_*=\eps_{pp} + \eps_{\rm{CNO}}$ is the nuclear energy
production per unit mass from both pp-chain and CNO cycle.

In order to get an expression for $dX_{\rm core}/dt$, we need to calculate the rate of change of the mass of hydrogen in the convective core. As shown in Appendix, we have

\begin{equation}
\frac{dX_{\rm core}}{dt}=-\frac{4 m_{\rm p} \overline{\eps}_*}{Q},
\label{Xbar}
\end{equation}
where $\overline{\eps}_*$ is the mass average of $\eps_*$, namely

\begin{equation}
\overline{\eps}_*= \frac{\int \eps_* dM}{M_{\text{core}}} = \frac{\int_{0}^{\pi} \int_{0}^{\eta_{\text{core}}}\eps_* \rho r_\zeta r^2 \sin\theta d\zeta d\theta}{\int_{0}^{\pi} \int_{0}^{\eta_{\text{core}}} \rho r_\zeta r^2 \sin\theta d\zeta d\theta}= \frac{L_{\rm core}}{M_{\rm core}} \ .
\end{equation}
$L_{\rm core}$ and $M_{\rm core}$ are respectively the luminosity and the mass of the convective core.  $\eta_{\rm core}$ is the core fractional radius, $(\zeta,\theta)$ are the spheroidal coordinates used
to describe the star (its surface is at $\zeta=1$) and $r_\zeta = \partial r/ \partial \zeta$ comes from the Jacobian associated with spheroidal coordinates \citep[e.g.][]{Rieutordal2016}. Finally, the discretised time evolution of $X_{\rm core}$ is given by

\begin{equation}\label{timeev}
X^{\rm core}_n=X^{\rm core}_{n-1} - \Delta t \frac{4 m_{\rm p}}{Q}\overline{\eps}_* \ ,
\end{equation}
where $n$ refers to the time-step of the computations. We now have a `clock' in the code allowing us to monitor the evolution
of $X_{\rm core}$ with time. In addition, we simplify the radial profile of the mean molecular weight at the outer boundary of the convective core by imposing a step-function.

\begin{figure}[t]
\includegraphics[width=.5\textwidth]{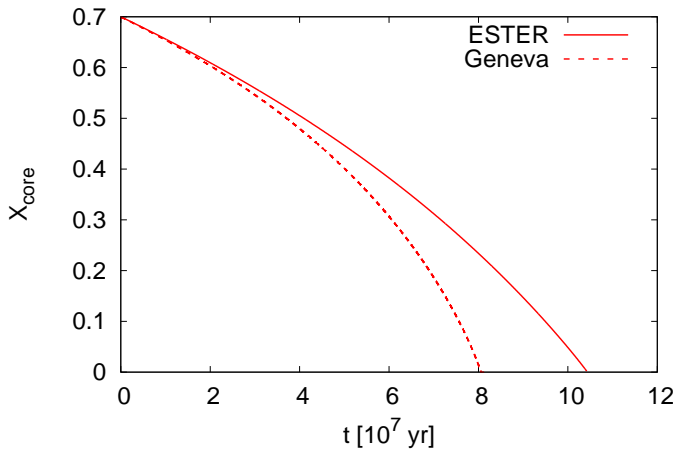}\hfill
\caption{Evolution of $X_{\text{core}}$ as a function of time for models of 5~$M_\odot$, $Z= 0.02$ \citep[GN composition,][]{GN93} computed with ESTER using the simple scheme for hydrogen burning (Eq. \ref{timeev}) and with  the Geneva code (Ekstr\"{o}m, private communication). These are non-rotating models computed with no mass loss and no overshooting.}
\label{fig:Xc_t}
\end{figure}

To check that our simplifications provide an acceptable evolution (especially in terms of lifetime on the MS), we compare the time variation of the core hydrogen mass fraction of the simplified ESTER set-up with the more realistic modelling used in the stellar evolution Geneva code. Fig.~\ref{fig:Xc_t} shows the evolution of $X_{\text{core}}/X_0$ at $Z=0.02$ with initial chemical mixture of \cite{GN93}, as a function of time for a non-rotating $5~M_\odot$ star computed with both the ESTER and the Geneva code (Ekstr\"{o}m, private communication), where $X_0$ is the initial hydrogen mass fraction on the zero age main sequence (ZAMS). The models are computed without convective overshoot (this is true for all models in the paper), and without mass loss. The result is that the time at which $X_{\rm core}$ vanishes, is slightly overestimated by the ESTER code. The difference is typically 20\%. We accept such a difference since it is not our purpose to give quantitative predictions, but to rather exhibit the main qualitative features of fast rotation evolution in  early-type stars. In the rest of the paper, we use $X_{\text{core}}/X_0$ as a proxy of time evolution on the MS. 


\subsection{The quasi-steady state approximation}

As mentioned above, the star can be considered in quasi-steady state if its relevant timescales are shorter than the timescales imposed by the drivers. Here, the relevant timescale is the Eddington-Sweet timescale $\Tau_{ES}$, which corresponds to
the time required for the redistribution of angular momentum (i.e. for baroclinic modes to be damped out, see \citealt{R06b}) and defined as

\begin{equation}
\Tau_{\rm ES}= \Tau_{\rm KH} \frac{GM}{\Omega^2_{eq} R^3}
\end{equation}
where $\Omega_{eq}$ is the equatorial angular velocity, and $\Tau_{\rm KH}=GM^2/(RL)$ is the Kelvin-Helmholtz
timescale. The comparison between the nuclear and mass loss timescales, $\Tau_{\rm nucl}$ and $\Tau_{\rm ml}$, and the Eddington-Sweet timescale at ZAMS for ESTER 2D-models of $5$, $10$ and $20~M_\odot$ stars initially rotating at $\omega_i= \Omega_{eq, i}/{\Omega_k}=0.3$ where $\Omega_k$ is the Keplerian angular velocity is given in Table~\ref{table:1}. 
It shows that the Eddington-Sweet timescale is at least one order of magnitude shorter
than the nuclear timescale and at least two orders of magnitude shorter than the mass loss timescale for our simulations, meaning that the
angular momentum has enough time to be redistributed during the stellar
evolution, and that the quasi-steady state is a good approximation.

\begin{table}[t]
\caption{Comparison between the Eddington-Sweet timescale $\Tau_{\rm ES}$,  the nuclear timescale $\Tau_{\rm nucl}$, and the mass loss timescale $\Tau_{\rm ml}$  for stars of different initial masses. These timescales are evaluated at ZAMS with an equatorial angular velocity such that $\Omega_{eq}=0.3~\Omega_k$. Metallicity is set to $Z=0.02$.}            
\label{table:1}      
\centering                                   
\begin{tabular}{c c c c}   
\hline 
\\        
$M/M_{\odot}$  & $\Tau_{\rm ES}$ (yr) & $ \Tau_{\rm nucl}$ (yr) & $ \Tau_{\rm ml}$ (yr)\\ 
\hline      \\                        
    5  & $6.4 \cdot 10^6$ & $1.1 \cdot 10^8$ & $1.78 \cdot 10^{11}$\\    
    \\                                     
    10  & $1.6 \cdot 10^6$ & $2.6 \cdot 10^7$ & $4.69 \cdot 10^{10}$\\    
    \\    
    20  & $5.9 \cdot 10^5$ & $9.6 \cdot 10^6$ & $9.1 \cdot 10^{8}$ \\    
    \\

\hline                                             
\end{tabular}
\end{table}

\section{Evolution at constant mass and angular momentum}\label{constant_angmom}


As a first step, we concentrate on `low-mass' early-type stars throughout
the Main Sequence (MS). Indeed, for such stars, we can neglect mass and angular momentum losses. This is a good approximation for $ M \lesssim 7M_\odot$ (this limit actually depends on the considered value of metallicity). 
We thus enforce constant angular momentum and constant mass, namely

\[ L_z=\intvol r^2\sin^2\theta\,\Omega \rho dV = {\rm Cst} \quad {\rm and}\quad M =
\intvol\rho dV = {\rm Cst}\ .  \] 
We compute 2D quasi-steady models with the ESTER code and follow the nuclear evolution according to equation (\ref{timeev}). 

\subsection{Evolution of surface stellar parameters}
\label{subsectionGlobal5Msun}

We first consider the main sequence evolution of surface stellar parameters of a $5~M_\odot$ model at two different metallicities $Z=0.02$ and $Z=10^{-3}$ (typical of Population II stars) and initially rotating at $50 \%$ of the critical angular velocity. Because such stars exhibit very weak radiative acceleration, their critical angular velocity is derived from the $\Omega$-limit, namely when the centrifugal acceleration balances gravity at equator (e.g. Paper I). In that case, the critical angular velocity is the Keplerian angular velocity $\Omega_k = \sqrt{GM/R^3_{eq}}$. 
 When this critical angular velocity is reached, the spin-up has to stop and the star loses angular momentum such that it remains below and near critical rotation. This angular momentum loss is assumed to occur through mechanical mass transfer into a decretion disc (e.g., Okazaki 2004, Krticka 2011).

\begin{figure}[t]
  \centerline{\includegraphics[width=.5\textwidth]{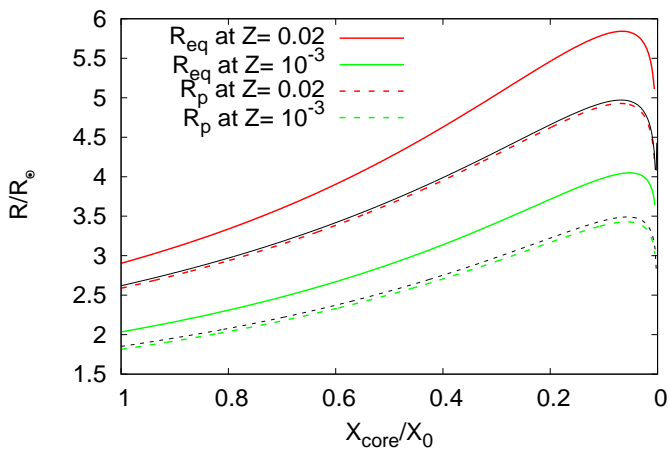}}
\caption{Equatorial and polar radius as a function of the fractional
abundance of hydrogen in the convective core for a $5~M_\odot$
star initially rotating at $\omega_i=0.5$ for $Z=0.02$ and
$Z=10^{-3}$. The stellar radius in case of no rotation has been added
for comparison in full and dashed black lines for $Z=0.02$ and $10^{-3}$ respectively.}
\label{fig:R_Xc}
\end{figure}
\begin{figure}[t]
\centerline{\includegraphics[width=.5\textwidth]{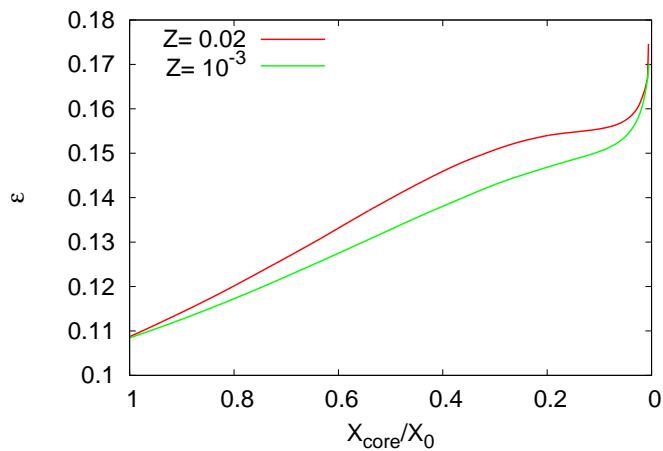}}
\caption{Same as Fig.~\ref{fig:R_Xc}, but for the surface flattening $\epsilon=1-R_{p}/R_{eq}$.}
\label{fig:epsilon_Xc}
\end{figure}
\begin{figure}[t]
  \centerline{\includegraphics[width=.5\textwidth]{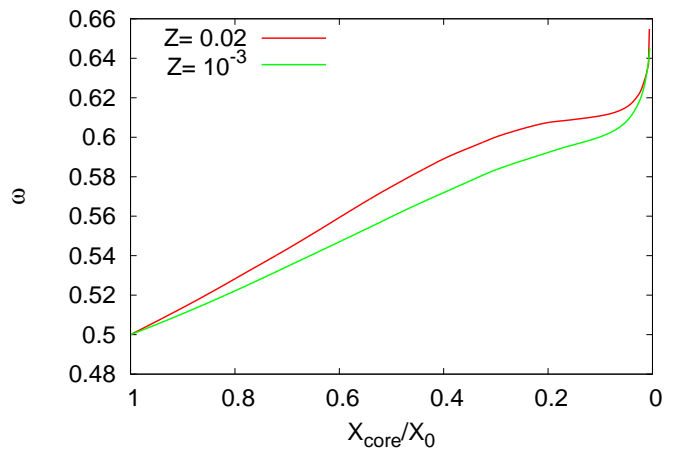}}
\caption{Same as Fig.~\ref{fig:R_Xc}, but for the angular velocity at the equator in units of the equatorial critical angular velocity $\omega=\Omega_{eq}/\Omega_k$.}
\label{fig:Omegabk_Xc}
\end{figure}
\begin{figure}[h]
\centerline{\includegraphics[width=.5\textwidth]{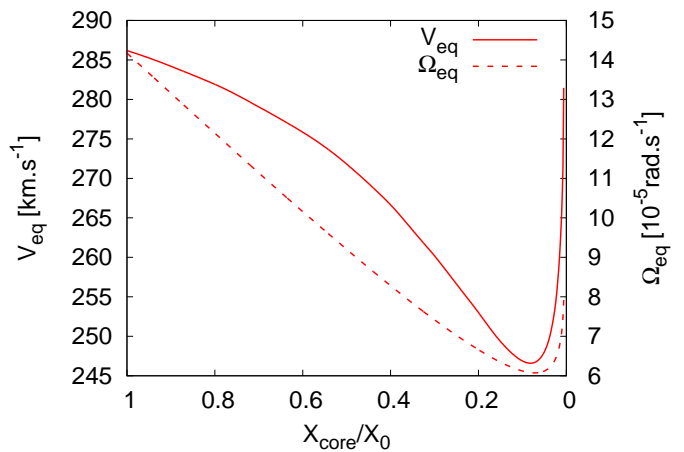}}
\caption{Same as Fig.~\ref{fig:R_Xc}, but for the equatorial velocity and angular velocity.}
\label{fig:Veq_Xc}
\end{figure}

 To describe the evolution of the rotational properties of the models, we focus on the evolution
of the equatorial and polar radii, $R_{eq}$ and $R_p$, the surface
flattening $\epsilon=1-R_{p}/R_{eq}$, the ratio of the equatorial
angular velocity to the Keplerian one $\omega$, and
the linear equatorial velocity $V_{eq}= R_{eq}\Omega_{eq}$. Figures \ref{fig:R_Xc}, \ref{fig:epsilon_Xc}, \ref{fig:Omegabk_Xc}, and \ref{fig:Veq_Xc} display the evolution of these quantities along the MS. 

\begin{figure}[t]
  \centerline{\includegraphics[width=.5\textwidth]{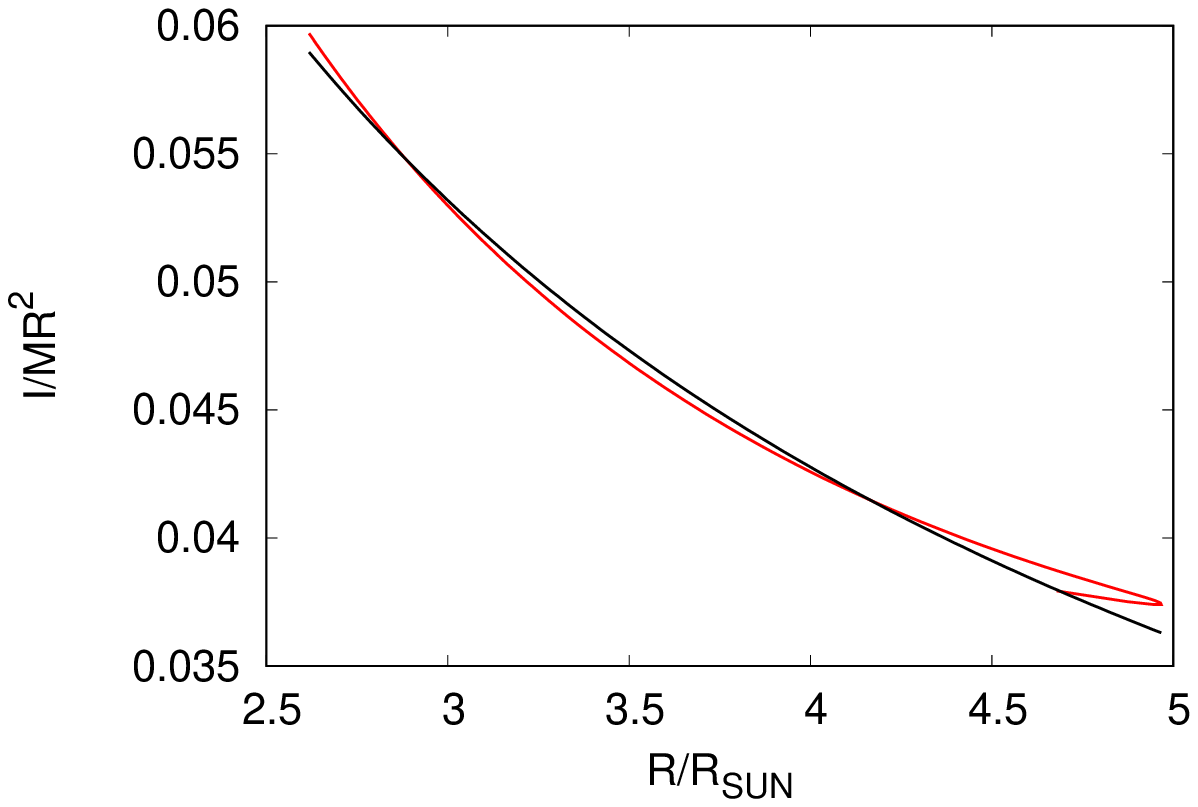}}
  \caption{MS evolution of the ratio $A=I/(MR^2)$ as a function of the stellar radius for a $5~M_\odot$ star initially rotating at $\omega_i= 0.1$ and with $Z= 0.02$. A rough fit represented with a black line gives $A \propto R^{-0.76}$.}
\label{fig:mom_inertia}
\end{figure}

\subsubsection{Stellar radius and surface flattening}
\label{subsubsectionGlobal5Msunflattening}

The mean stellar radius is well known to increase along the MS, the model with the lowest Z being more compact \citep[e.g.][]{Yoon2006, Ekstrom2011}. Fig.~\ref{fig:R_Xc} shows that rotating stars expand faster at the
equator than at the pole. Indeed, the centrifugal effect affects only weakly the
polar regions, and does not contribute to the polar radius growth; however, it
does affect the equator significantly.  The different behaviour of polar and equatorial radii can be seen in Fig.~\ref{fig:epsilon_Xc} where the flattening  of the star $\epsilon$ is plotted along the main sequence evolution. The rapid
expansion of the equatorial radius yields a rapid growth of $\epsilon$ for the two values of Z. The stellar expansion due to nuclear evolution  is slower for the less metallic star, since its keplerian angular velocity $\Omega_k$ decreases more slowly, leading to a slower growth of $\omega$ and thus of $\epsilon$.

\subsubsection{ Surface rotation}
\label{subsubsectionSurfacerotation}

Fig.~\ref{fig:Omegabk_Xc} shows the evolution of the angular velocity ratio $\omega=\Omega_{eq}/\Omega_k$ throughout the MS. The fact that it increases as $X_{\rm core}/X_0$ decreases can
be easily explained. Indeed, assuming slow rotation, thus weak flattening and  weak differential rotation, we can estimate the moment of inertia of the star as

\begin{equation}
I=\int r^2 dm=AMR^2,
\end{equation}
where $A$ informs us on the stellar mass distribution. The smaller $A$, the more centrally condensed the star. Fig.~\ref{fig:mom_inertia} shows the evolution of $A=I/MR^2$ as a function of the stellar radius along the MS for a $5~M_\odot$ star initially rotating at $\omega_i=0.1$ and with $Z=0.02$. For this model, a rough fit shows that $A$ decreases as $\sim R^{-0.76}$. Therefore, as the star evolves along the MS, it becomes more and more centrally condensed. At constant angular momentum $L= AMR^2 \Omega = \rm Cst$, and as the star expands,  $\Omega$ decreases as $\propto R^{-1.24}$. $\Omega$ thus decreases more slowly than $\Omega_k \propto R^{-1.5}$,  leading to an increased angular velocity ratio $\omega$ throughout the MS.
The foregoing remark shows that the natural trend of the evolution of rotation is an increase of $\omega$, namely an increase of criticality along the MS. The full calculation with 2D models confirms this behaviour at high rotation rate (Fig.~\ref{fig:Omegabk_Xc}). Metallicity also plays a role in the evolution of $\omega$: as mentioned before, a less metallic star expands more slowly (see Fig.~\ref{fig:R_Xc}) and thus $\Omega_k$ decreases less rapidly, leading to a slower increase of $\omega$.
We stress that the star does not
rotate faster as it evolves through the MS even though $\omega$ increases. Actually, it is the opposite up to the so-called hook at the end of the MS, as shown by Fig.~\ref{fig:Veq_Xc}. After the hook,  nuclear heat generation no longer compensates the radiated energy leading the star to an overall contraction \citep{iben67}. The conservation of angular momentum then leads to an increase of equatorial angular velocity $\Omega_{eq}$.

\subsubsection{Gravity darkening}
\label{subsubsectionGravitydarkening}

\begin{figure}[t]
\includegraphics[width=.5\textwidth]{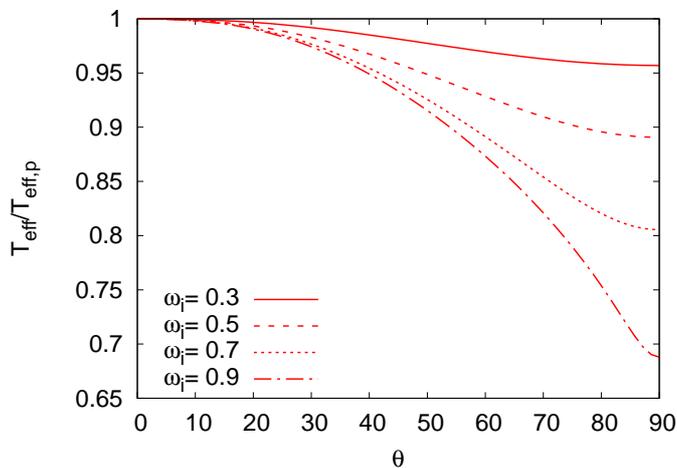}
 \caption{Variation of the effective temperature at ZAMS as a function of colatitude for a $5~M_\odot$ star at $Z=0.02$ for various values of $\omega_i$. }
\label{fig:Teff_theta}
\end{figure}

The flattening of rapidly rotating stars  gives rise to gravity darkening, which reduces the flux at the equator \citep[][]{VonZeipel1924}. 
Fig.~\ref{fig:Teff_theta} shows the
variation of the ratio between the local effective temperature and the polar effective temperature  as a function of colatitude for the
$5~M_\odot$ model with $Z=0.02$, at ZAMS and for various values of $\omega_i$. The latitudinal variation of this ratio at $Z=10^{-3}$ is not shown because almost identical to the case with $Z=0.02$, with a relative difference never exceeding $0.3\%$. The effective temperature drop between pole and equator is thus almost exactly the same for the two metallicities, as expected from Fig.~\ref{fig:epsilon_Xc} showing that both models have the same surface flattening at ZAMS.

\subsection{Initial angular velocity requirements for criticality}

To appreciate the initial states (at ZAMS) 
that lead to the critical velocity before the end of the MS,
we compute a grid of models with constant masses $5 \leq M/M_{\odot} \leq 10$ and  with $Z=0.02$ and $Z=10^{-3}$,
backward in time. Hence we start with models rotating near criticality
(here we take $\omega=0.9$, i.e. quasi-criticality) at various hydrogen mass fractions in the core ($X_{\rm core,c}/X_0$), or equivalently at various stage along the MS, and increase the hydrogen mass fraction until $X_{\rm core}=X_0$, namely until ZAMS is reached. 

\begin{figure}[t]
  \centerline{\includegraphics[width=.5\textwidth]{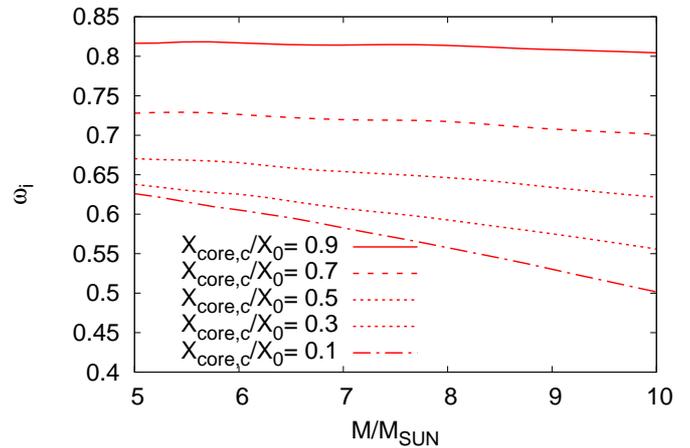}}
  \caption{Values of the ZAMS angular velocity ratio $\omega_i$ required to reach quasi-criticality ($\omega= 0.9$) at different values of the hydrogen mass fraction in the core, as a function of the stellar mass $M/M_{\odot}$ for $Z=0.02$, and without any mass loss. These predictions result from computations backward in time.
}
\label{fig:Omega_M_nomloss}
\end{figure}

Figure~\ref{fig:Omega_M_nomloss} shows the ZAMS angular velocity ratio $\omega_i$ needed to reach quasi-criticality during the MS for various hydrogen mass fractions in the core at quasi-criticality, as a function of the stellar mass $M/M_\odot$, when $Z=0.02$. 
The  more massive the star, the smaller $\omega_{i}$ is needed to reach critical rotation on the MS, in particular for $X_{\rm core}/X_0 \leq 0.5$. Moreover, the smaller the fractional
abundance of hydrogen left in the convective core at quasi-criticality,
the smaller $\omega_{i}$ needs to be. Indeed, as mentioned before, stellar expansion tends to make the angular velocity ratio $\omega$ increase as the star evolves through the MS. Therefore, the smaller the fractional
abundance of hydrogen left in the convective core at quasi-criticality, the longer the evolution along the MS and thus the more $\omega$ has increased.
At ZAMS, the star therefore does not need to be rotating very rapidly  to reach criticality near the end of the MS.

\begin{figure}[t]
  \centerline{\includegraphics[width=.5\textwidth]{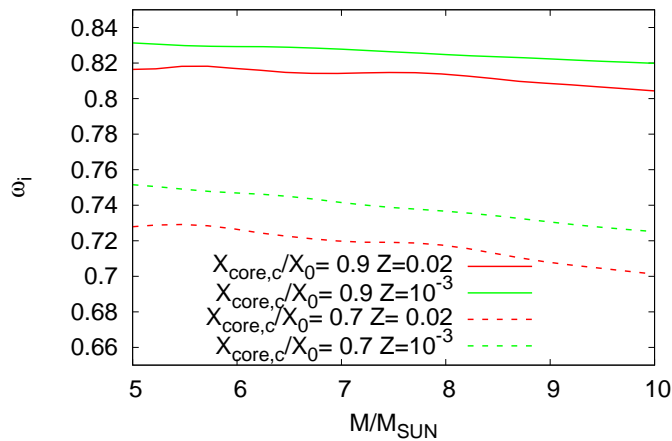}}
  \caption{Same as Fig.~\ref{fig:Omega_M_nomloss} for $X_{\rm core,c}/X_0= 0.9$ and $0.7$, and for $Z=0.02$ and $10^{-3}$.
}
\label{fig:Omega_M_nomloss_Z}
\end{figure}

We now take a look at the effect of lower metallicity on the initial conditions required to reach criticality before the end of the MS.  
The comparison between the ZAMS angular velocities at equator in units of the equatorial critical angular velocity as a function of the mass for $Z=0.02$ and $Z=10^{-3}$ is shown in Fig.~\ref{fig:Omega_M_nomloss_Z}. 
We clearly see that for a lower metallicity, the value of the initial
angular velocity required to reach criticality at a given hydrogen content on the MS is higher than at higher metallicity. This reflects the slower expansion of less metallic stars.


\section{Evolution with angular momentum loss}
\label{sectionEvolutionwithAMloss}

We now consider stars more massive than $\sim 7~M_\odot$ for which mass loss from radiation-driven winds can no longer be neglected.

\subsection{The critical angular velocity at the $\Omega\Gamma$-limit}

The critical angular velocity of massive stars is expected to differ from the Keplerian angular velocity because of radiative acceleration (see  \citealt{MaederMeynet2000}). Using the $\theta$-dependent radiative flux expression from the $\omega$-model of \cite{ELR11}, we have shown in Paper I that, in this approximation, the angular velocity becomes critical when the rotation-dependent Eddington parameter at equator

\begin{equation}
    \Gamma_{\Omega}(\pi/2)= \Gamma_{eq} \left(1-\frac{\Omega^2 R_{eq}^3}{GM}\right)^{-2/3} \ ,
\end{equation}
reaches unity, where

\begin{equation}
\Gamma_{eq} = \frac{\kappa(\pi/2)L}{4 \pi c G M} \ ,
\end{equation}
is the standard Eddington parameter at equator. The critical angular velocity then reads

\begin{equation}\label{eq:omegac}
\Omega_c=\Omega_k \sqrt{1-\Gamma_{eq}^{3/2}}
\; .
\end{equation}

\begin{figure}[t]
  \centerline{\includegraphics[width=.5\textwidth]{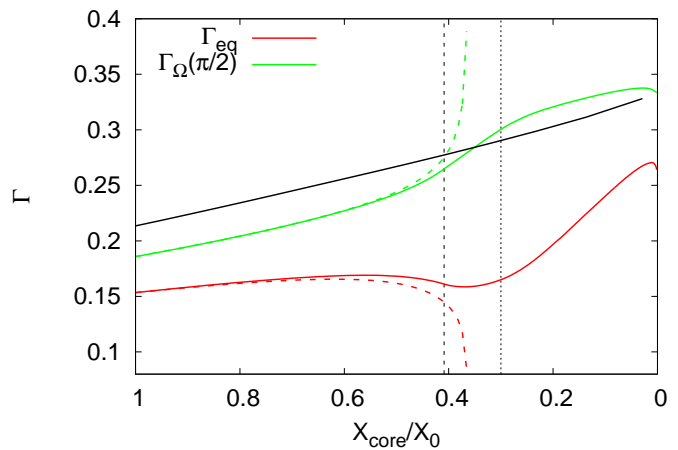}}
  \caption{$\Gamma_{eq}$ and $\Gamma_{\Omega}(\pi/2)$ as a function of the fractional abundance of hydrogen in the convective core for a $15~M_\odot$ ESTER 2D-model initially rotating at $\omega_i=0.5$, with and without mass loss (respectively solid and dashed lines). The  black line shows the evolution of $\Gamma_{eq}$ for $\omega=0$. The two vertical lines delineate the two phases of the two-wind regime (see sect.\ref{sec:M15Om05}).}
\label{fig:Gamma_eq}
\end{figure}

Thus, it is expected that the critical angular velocity is reduced by radiative acceleration at the stellar surface. However, as shown in Paper I, this reduction is actually quite small because of the combined effects of gravity darkening and opacity reduction at equator. We further illustrate this property in Fig.~\ref{fig:Gamma_eq} where we plot the evolution of $\Gamma_{eq}$ and $\Gamma_{\Omega}(\pi/2)$ for a 15~\msun\ ESTER 2D-model. Unlike the no-mass loss case discussed in paper I, mass loss prevents the star from reaching criticality (see below) making  $\Gamma_{\Omega}(\pi/2)$ always finite. For such a star, $\Gamma_{eq}\infapp0.27$, implying that $\Omega_c/\Omega_k\supapp0.93$ all the time. This result shows that the difference between critical and Keplerian angular velocity remains small. Thus, we shall continue to express $\Omega_{eq}$ as a fraction of the equatorial Keplerian angular velocity $\Omega_k$ to measure the distance to criticality.

\subsection{Mass and angular momentum loss}\label{sec:massandAMloss}

Similarly to the case of evolution without angular momentum loss, we evolve the star with ESTER through the MS by decreasing the fractional abundance of hydrogen in the convective core $X_{\rm core}/X_0$. However, at each step, we also remove some mass and angular momentum off the star, for that we use the clock given by the time evolution of $X_{\rm core}$ in Eq.~(\ref{Xbar}).

We estimate the mass-loss rate using the local mass-flux prescription of Paper I, which can be seen as a local equivalent of the modified CAK theory \citep{PPK} for rotating winds. Although not clearly assessed by observations (see the discussion in Paper I), we also include the bi-stability jump  in our local mass loss modelling.

The total mass and angular momentum loss rates for a rotating star read

 \begin{equation}
\dot{M} =  \int \dot{m}(\theta) dS \qquad {\rm and}\qquad
\dot{\mathcal{L}} = \int \dot{\ell}(\theta) dS  \ .
\end{equation}
The area element at the stellar surface is

 \begin{equation*}
 dS=2 \pi R^2(\theta)\sqrt{1+\frac{R^2_{\theta}}{R^2({\theta)}}}\sin\theta d\theta \ ,
 \end{equation*}
 where $R(\theta)$ is the $\theta$-dependent radius of the star and  $R_{\theta}=\partial R/\partial \theta$ \citep{Rieutordal2016}. From Paper I, the local mass-flux is estimated via

\begin{equation}\label{eq:m1}
\begin{aligned}
\dot{m}(\theta) =&\; \frac{4}{9} \frac{\alpha(\theta) k'(\theta)^{1/\alpha'(\theta)}}{ v_{\rm th}(\theta) c} \\
\times &\; \left[ \frac{c}{\kappa_e (1-\alpha)}\left(|g_{\rm eff}(\theta)|- \frac{\kappa_e F(\theta)}{c}\right)\right]^{\frac{\alpha'(\theta)-1}{\alpha'(\theta)}} \\
\times &\;  F(\theta)^{1/\alpha'(\theta)}\ ,
\end{aligned}
\end{equation}
and the local angular momentum flux reads

\begin{equation}
\dot{\ell}(\theta)=\dot{m}(\theta) \Omega(\theta) R({\theta)}^2\sin^2\theta \ ,
\end{equation}
where $\Omega(\theta)$ is the $\theta$-dependent surface angular velocity. In Eq.~(\ref{eq:m1}), $g_{\rm eff}(\theta)$ is the local gravito-centrifugal acceleration, $F(\theta)$ is the local radiative flux, $\kappa_e$ is the mass absorption coefficient for electron scattering, and $v_{\rm th}\equiv (2k_{\rm B}T_{\rm eff}/m_{\rm Fe})^{1/2}$ is the local iron thermal velocity. $\alpha$ and $k$ are the CAK force multiplier parameters (FMPs), and $\alpha'=\alpha - \delta$ where $\delta$ is another FMP related to the effect of radial changes of
ionisation when moving outward in the wind. We assume $\delta$ not to vary on the stellar surface, and equal to $0.1$, a typical value for hot stars at metallicity close to solar \citep{Abbott1982}. We also use the parametrisation of $\alpha$ and $k$ at $Z= 0.02$ from Paper I, namely

\begin{equation}\label{alpha}
\alpha(T_{\text{eff}}) =
\begin{cases}
    0.45, & \text{if $T_{\text{eff}} \leq 10\rm{kK}$ \ ,}\\
    1.5\times 10^{-5} T_{\rm eff} + 0.3, & \text{if $10\rm{kK} <  \Teff \leq 20~{\rm kK}$ \ ,}\\
        5\times 10^{-6} T_{\rm eff} + 0.5, & \text{if $20\rm{kK} <  T_{\text{eff}} \leq 40~{\rm kK}$ \ ,}\\
            0.7, & \text{if $T_{\text{eff}} > 40~{\rm kK}$ \ ,}\\
  \end{cases}
\end{equation}
and

\begin{equation}\label{k}
\begin{aligned}
&k(T_{\rm eff}) \simeq \\
&\begin{cases}
    \exp(-2.15 \times 10^{-4}  T_{\rm{eff}} + 2.41), & \text{if $T_{\text{eff}} \leq 20~{\rm kK}$,}\\
    -3.00 \times 10^{-6}  T_{\rm{eff}} + 0.22, & \text{if $20~{\rm kK} <T_{\text{eff}} \leq T_{\text{eff}}^{\rm jump}$,}\\
        1.16 \times 10^{-6} T_{\rm{eff}} + 0.08, & \text{if $T_{\text{eff}} > T_{\text{eff}}^{\rm jump}$.}\\
  \end{cases}
  \end{aligned}
\end{equation}
$T_{\rm eff}^{\rm jump}$ is the effective temperature of the bi-stability jump, that is, the temperature below which Fe IV recombines into Fe III, and is defined as \citep{Vink2001}

\begin{equation}
T_{\text{eff}}^{\text{jump}} = 61.2+ 2.59\moy{\rho}  \ ,
\end{equation}
where $\moy{\rho}$ is the characteristic wind density at 50\% of the terminal velocity of the wind. It is given by 
\begin{equation}
\log\moy{\rho} = -14.94  + 3.2 \Gamma_e \ ,
\end{equation}
where
\begin{equation}
  \Gamma_e = \frac{\kappa_e L}{4 \pi c G M}   \ ,
\end{equation}
is the classical Eddington parameter.
In 1D, the Fe IV--Fe III recombination theoretically leads to an increased  global mass-loss rate and a decreased terminal velocity of the wind. The effective temperature at the surface of rotating stars is however $\theta$-dependent and the bi-stability jump can therefore occur locally if there is a latitude on the stellar surface where $T_{\rm eff} = \Tej$.
We note that (\ref{k}) has been obtained assuming that the global mass-loss rate in the non-rotating regime $\dot{M}= 4 \pi R^2 \dot{m}$ is equivalent to the one of \cite{Vink2001} (some caveats on the use of this mass loss prescription for the calibration of $k$ are discussed in Paper I). $k$ has been calibrated with ESTER models at ZAMS, although it slightly varies with the evolutionary stage along the MS due to the evolution of $g_{\rm eff}$. We find that $\Delta k = k(X_{\rm core}/ X_0=0.1) - k(X_{\rm core}/ X_0=1)> 0 $ never exceeds 20\% of $k(X_{\rm core}/ X_0=1)$ for all $T_{\rm eff}$. Since our minimum value for $\alpha'$ is $0.35$ and $\dot{m} \propto k^{1/\alpha'}$, the local mass-flux may be underestimated by a factor $\sim 1.7$, at most. This difference is  acceptable considering the other approximations used and the uncertainties of the wind model.

\begin{figure}[t]
  \centerline{\includegraphics[width=.5\textwidth]{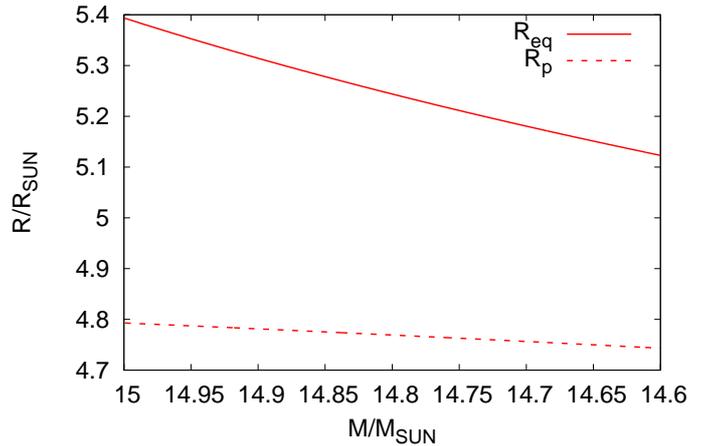}}
  \caption{Equatorial and polar radius as a function of the mass for a $15~M_\odot$ star with $Z=0.02$ initially rotating with $\omega_{i}=0.5$  and losing a total amount of mass of $\Delta M \simeq 0.4 M_\odot$.}
\label{fig:R_it_mloss}
\end{figure}

\subsection{A remark on mass loss effects without nuclear evolution}
\label{sec:nonuc}

At this stage, it is useful to recall how the radius of a star changes when a small amount of mass is removed. The outcome is well-known for the case of fully convective stars, which are well represented by a $n=1.5$ polytrope. For such stars, if a small amount of mass is removed from the surface, the remaining mass expands \citep{Chandrasekhar1967}. On the other hand, radiative stars, which are well represented by $n>3$ polytropes, contract when subject to mass loss \citep{Heisler1986}. This is also true for ESTER 2D-models of massive stars with a radiative envelope, as can be seen in Fig.~\ref{fig:R_it_mloss} showing the evolution of the polar and equatorial radius of a $15~M_\odot$ star initially rotating at  $\omega_{i}=0.5$, from which we have extracted a total mass of $\Delta M \simeq 0.4~M_\odot$, keeping $X_{\rm core}/X_0=1$ constant. 

\begin{figure}[t]
  \centerline{\includegraphics[width=.5\textwidth]{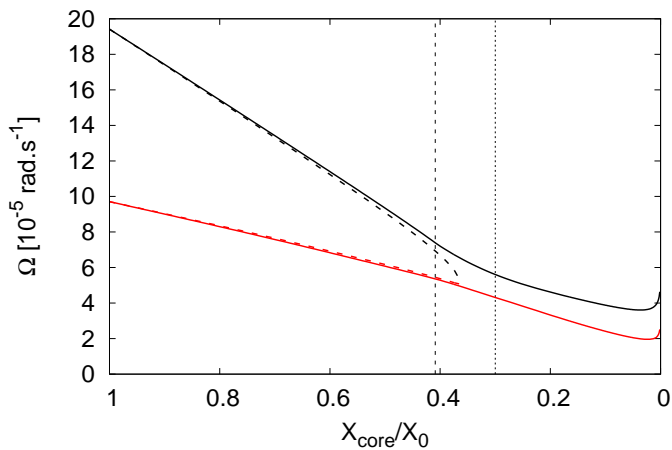}}
  \caption{Angular velocity (red) and equatorial Keplerian angular velocity (black) as a function of the fractional abundance of hydrogen in the convective core $X_{\rm core}/X_0$ for a $15 M_\odot$ star with a ZAMS angular velocity ratio $\omega_{i}=0.5$, with and without mass loss (respectively solid and dashed lines).}
\label{fig:Omegabk_Xc__}
\end{figure}

Unfortunately, the interpretation of the $\omega$-dependence on mass and angular momentum loss is rather complicated due to the non-linearity of the processes involved. While the angular momentum loss of massive stars only affects the evolution of $\omega$ by strengthening the decrease of $\Omega_{eq}$, the effect of slower expansion due to mass loss is twofold, since it leads to a slower decrease of both $\Omega_k$ and $\Omega_{eq}$. Whether mass loss itself tends to accelerate or slow down the natural trend of $\omega$ to increase throughout the MS is therefore surely model dependent. Fig.~\ref{fig:Omegabk_Xc__} shows the evolution of both $\Omega_{eq}$ and $\Omega_k$ with and without mass and angular momentum loss, for a 15~$M_\odot$ ESTER 2D-model with $Z=0.02$ and for $\omega_i=0.5$. For this specific model, we see that the evolution of $\omega$ almost only depends on the evolution of $\Omega_k$, that is because the effect of mass loss and angular momentum loss on the evolution of $\Omega_{eq}$ roughly compensate.

\subsection{The two different wind regimes}
\label{2windregime}

As shown in Paper I, the radiation-driven wind of rotating massive stars can be in two distinct regimes. The first one is a single-wind regime (SWR), characterised by a maximum mass-flux at the poles, which is effective when the effective temperature of the star is either greater or smaller than $\Tej$ for all $\theta$. The second one is a two-wind regime (TWR), which appears when there exists a colatitude $\theta_{\rm jump}$ on the stellar surface where $T_{\rm eff}(\theta_{\rm jump}) = \Tej$. The latter regime is characterised by a maximum mass-flux at equator and a stronger global mass and angular momentum loss rates. Intermediate-mass stars (or well-evolved massive stars, i.e. late B- and A-supergiants) can be in a cold-sided SWR, that is, with  $T_{\rm eff}< \Tej$ for all $\theta$, characterised by stronger global mass and angular momentum loss rates than in a TWR. Finding the general rules that govern the evolution of rotating massive stars turned out to be quite cumbersome since numerous particular cases pop up. Hence, to get an idea of the rotational evolution of massive stars, we focus on the case of a typical 15~M$_\odot$ star initially rotating at 50\% of the Keplerian angular velocity.

\subsection{The example of a $15~M_\odot$ 2D-model with $\omega_i=0.5$}\label{sec:M15Om05}

\begin{figure}[t]
  \centerline{\includegraphics[width=.5\textwidth]{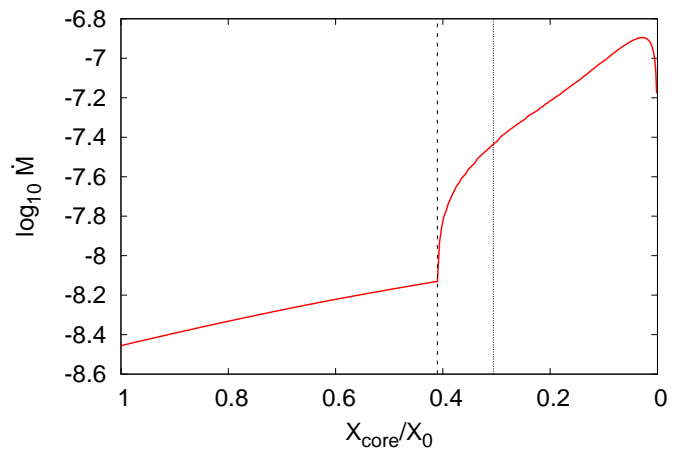}}
 \centerline{\includegraphics[width=.5\textwidth]{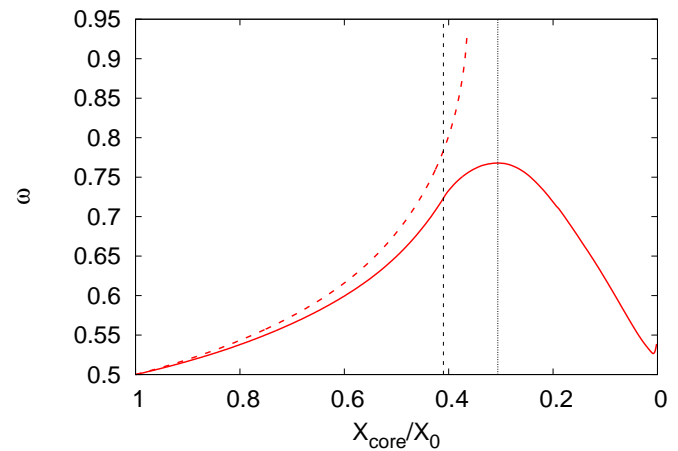}}
  \caption{Top: evolution of the mass-loss rate $\dot{M}$ (in $M_\odot \cdot  {\rm yr}^{-1}$) as a function  of the fractional abundance of hydrogen in the convective core $X_{\rm core}/X_0$ for a $15~M_\odot$ star with $Z = 0.02$ and for $\omega_i=0.5$. The vertical dashed line marks the evolutionary time at which the star reaches a TWR. The vertical dotted line marks the transition between Phase 1 and 2. Bottom: same but for the angular velocity at equator in units of the equatorial critical angular velocity $\omega=\Omega_{eq}/\Omega_k$. The red dashed line corresponds to the evolution of $\omega$ with constant mass and angular momentum throughout the MS.}
\label{fig:mdot_Xc}
\end{figure}

\begin{figure}[t]
     \centerline{\includegraphics[width=.5\textwidth]{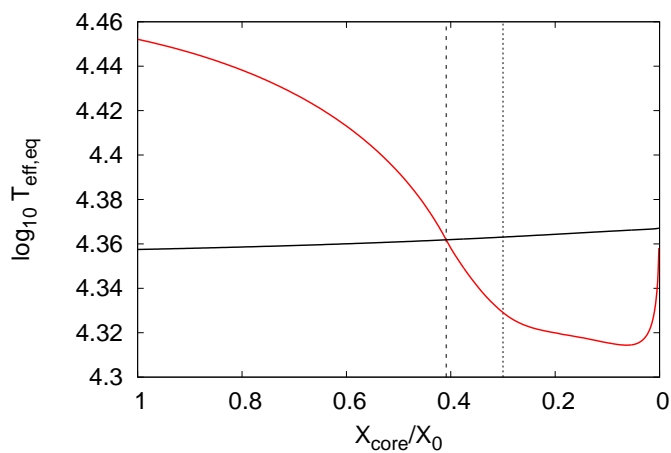}}
  \caption{As Fig.~\ref{fig:mdot_Xc}, but for the evolution of the equatorial effective  temperature $T_{\rm eff, eq}$ (in $\rm K$) as a function  of the fractional abundance of hydrogen in the convective core $X_{\rm core}/X_0$ for a $15~M_\odot$ star with $Z = 0.02$ and for $\omega_i=0.5$. The black full line corresponds to the evolution of the effective temperature of the bi-stability jump $T_{\rm eff}^{\rm jump}$.}
\label{fig:Teffeq_Xc}
\end{figure}

\begin{figure}[t]
\centerline{\includegraphics[width=.5\textwidth]{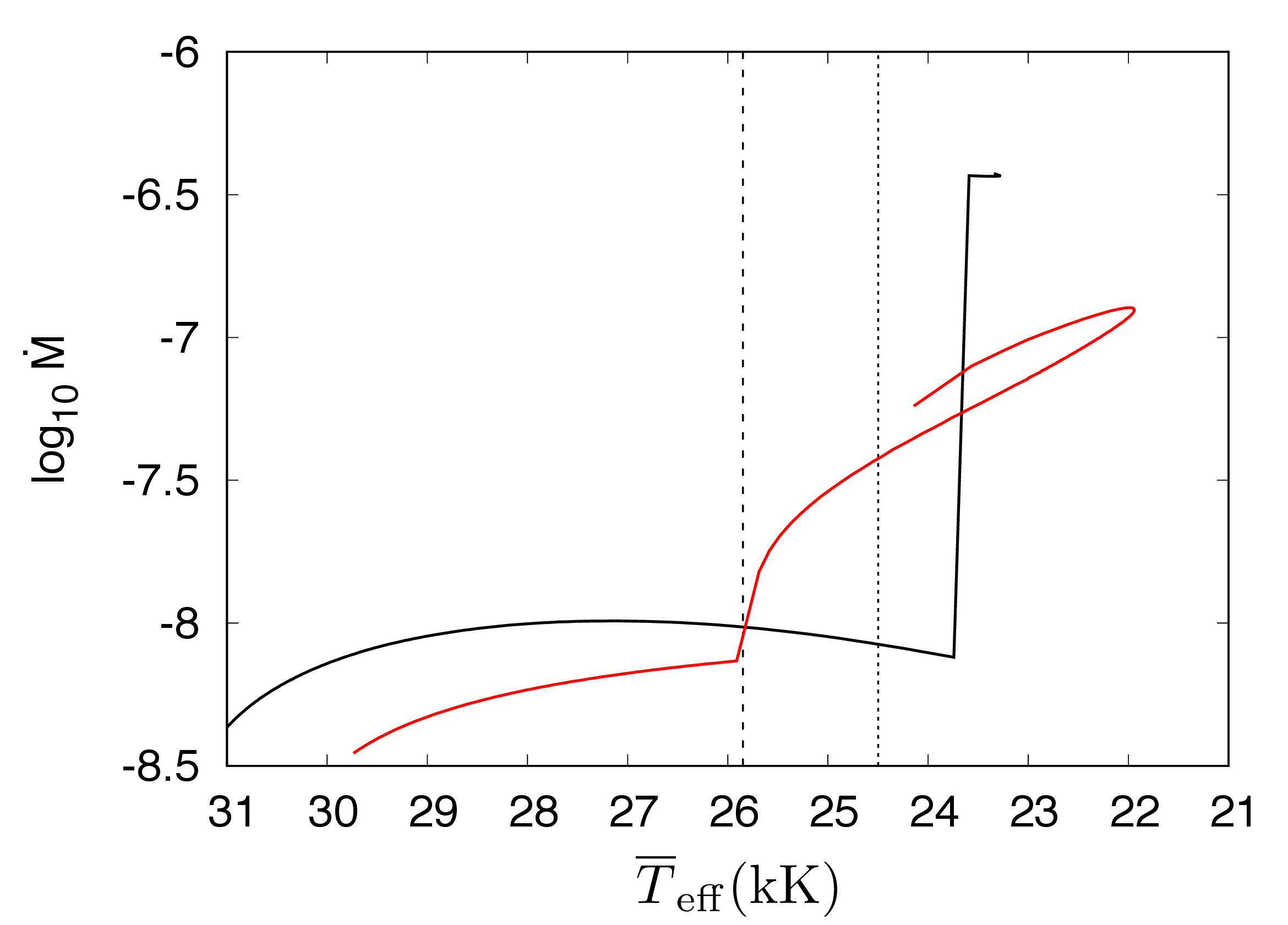}}
 \caption{Evolution of the mass-loss rate $\dot{M}$ as a function of the mean effective temperature $ \overline{T}_{\rm eff}$ for a $15~M_\odot$ ESTER 2D-model with $Z = 0.02$ and for $\omega_i=0.5$. The star evolves from left to right with a mean effective temperature approaching $30{\rm kK}$ at ZAMS, and $\overline{T}_{\rm eff} \simeq 22{\rm kK}$ at TAMS. The black curve corresponds to case of a non-rotating $15~M_\odot$ star with $Z = 0.02$ for which the mass-loss rate has been calculated using \cite{Vink2001} prescription. The vertical dashed line marks the evolutionary time at which the star reaches a TWR. The vertical dotted line marks the transition between Phase 1 and 2.}
 \label{fig:mdot_Teff}
\end{figure}

\begin{figure}[t]
	\includegraphics[width=0.24\textwidth]{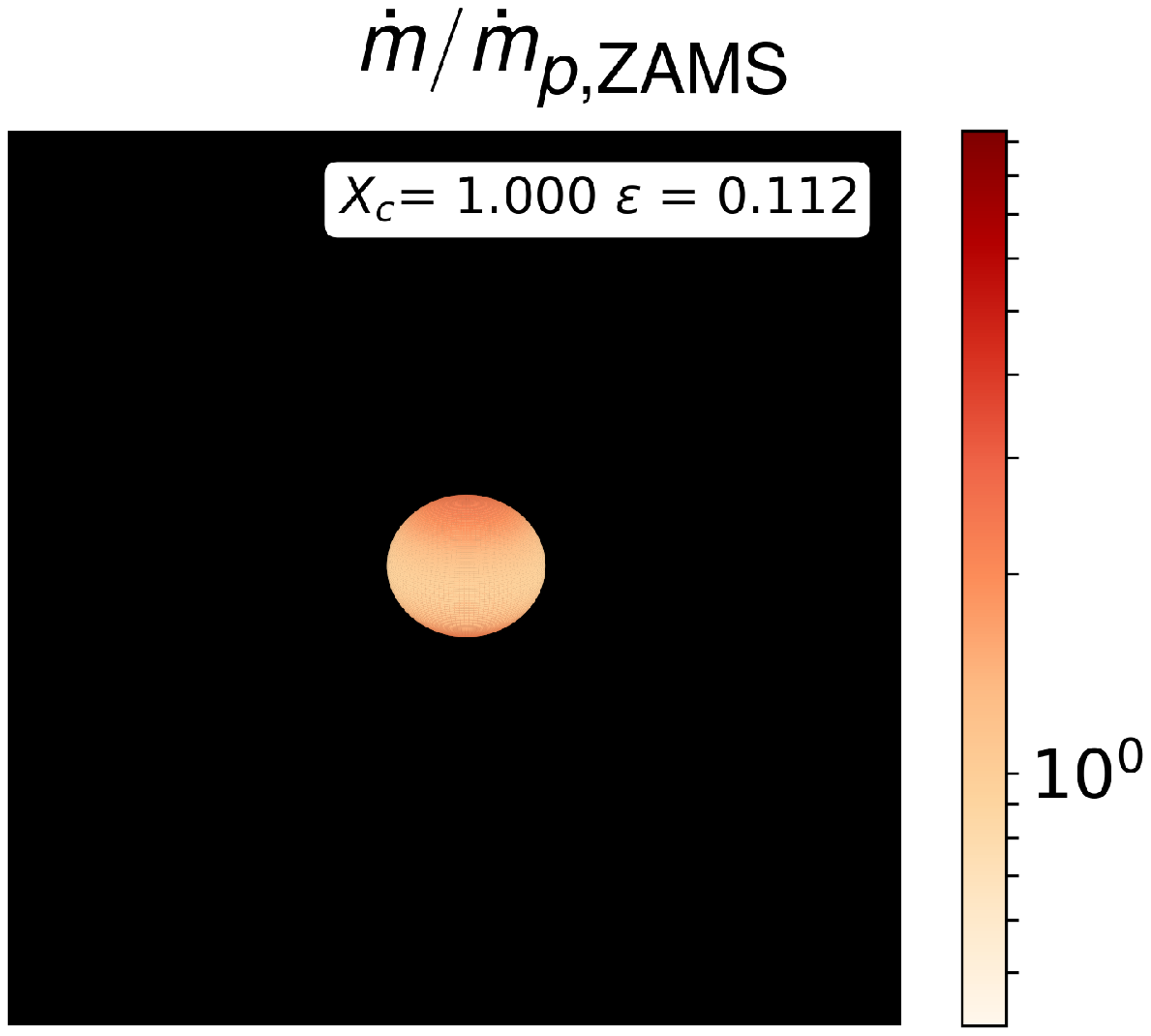}\hfill
      \includegraphics[width=0.24\textwidth]{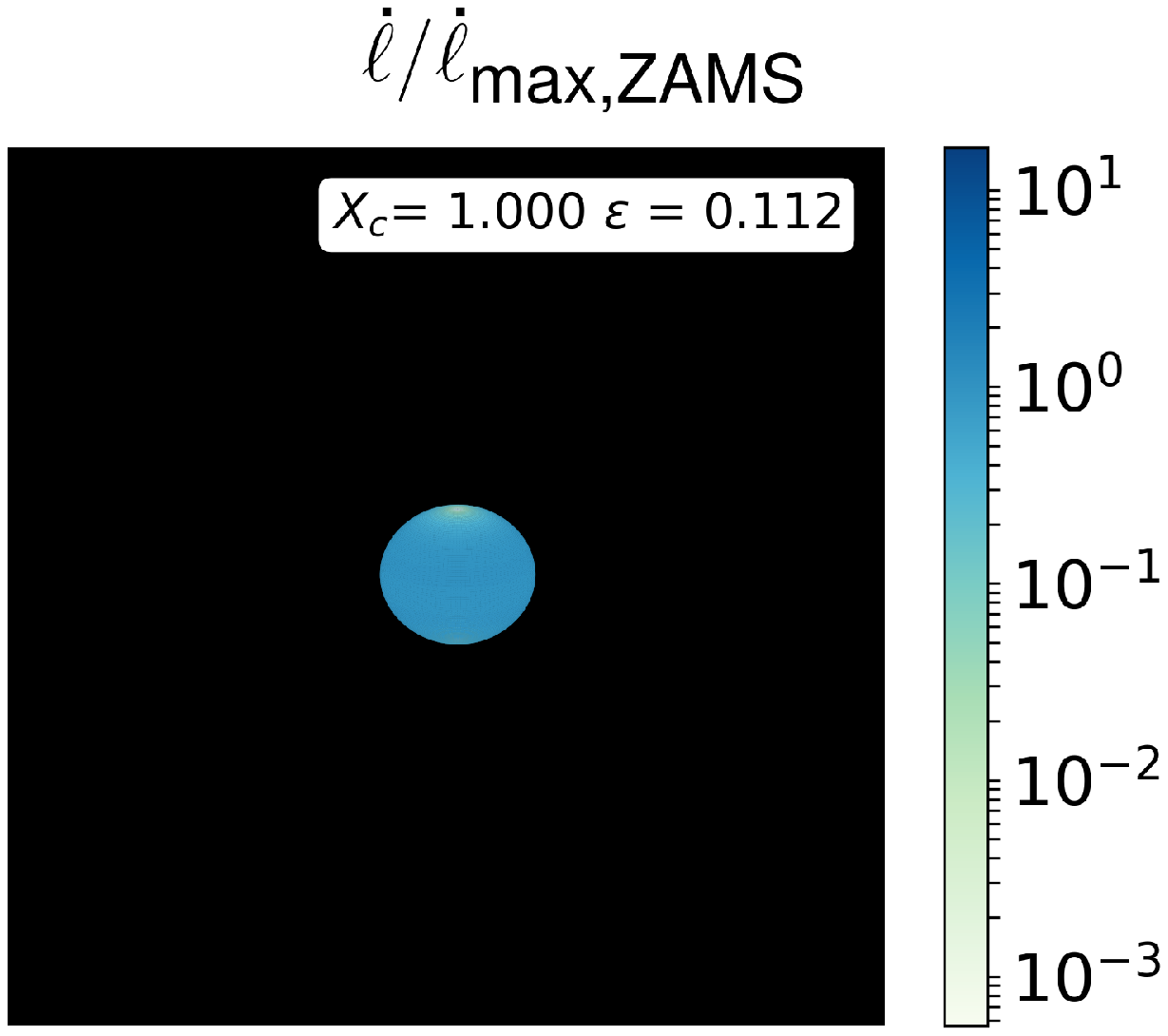}\\
	\includegraphics[width=0.24\textwidth]{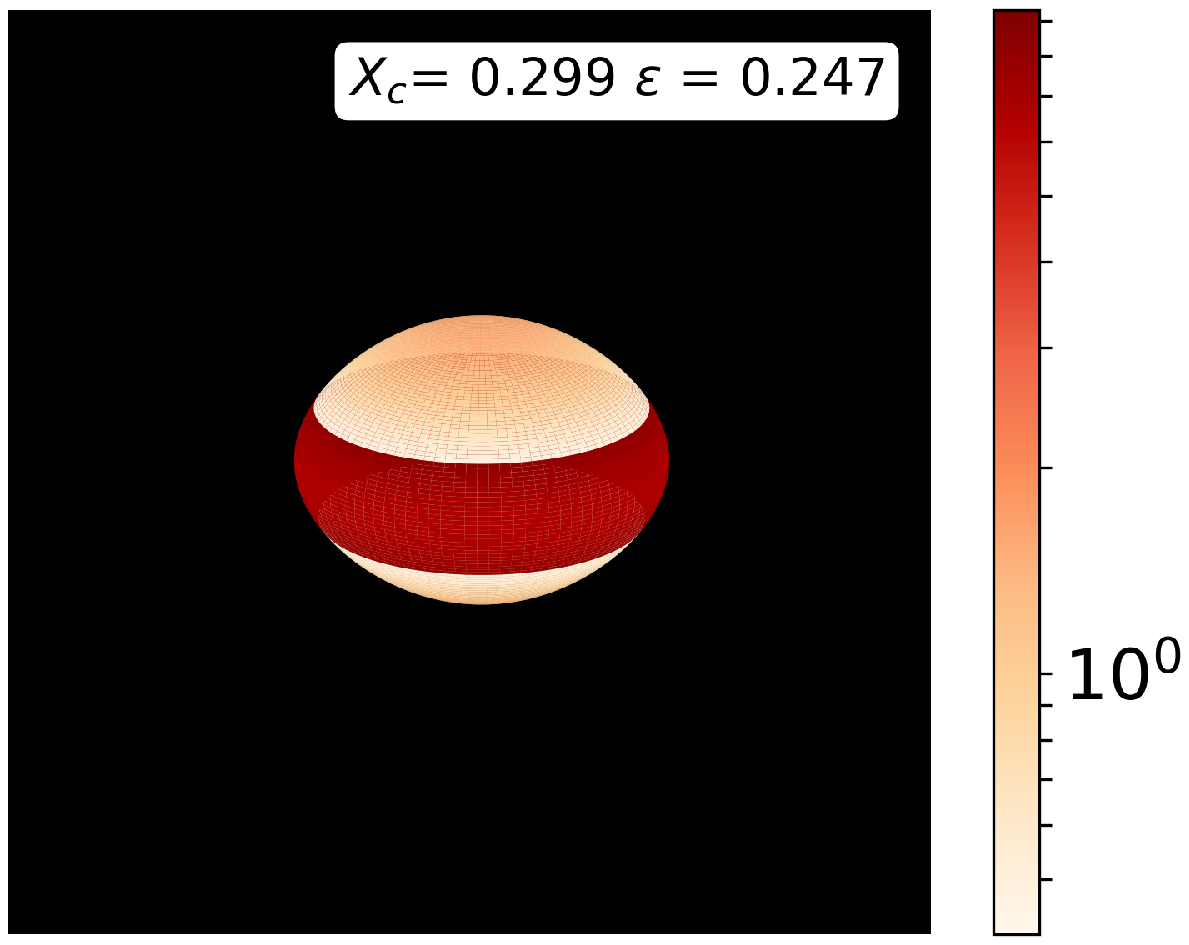}\hfill
    	\includegraphics[width=0.24\textwidth]{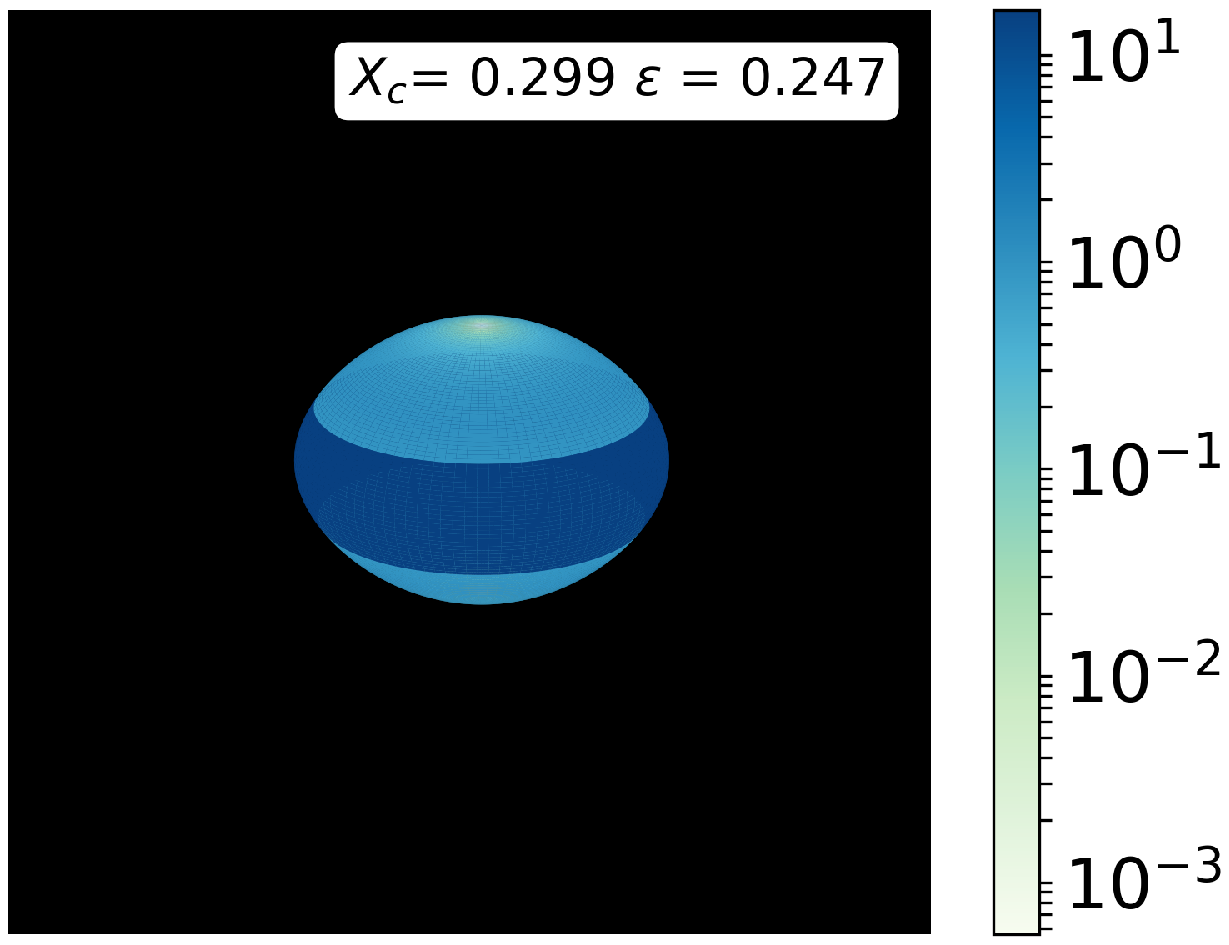}\\
	\includegraphics[width=0.24\textwidth]{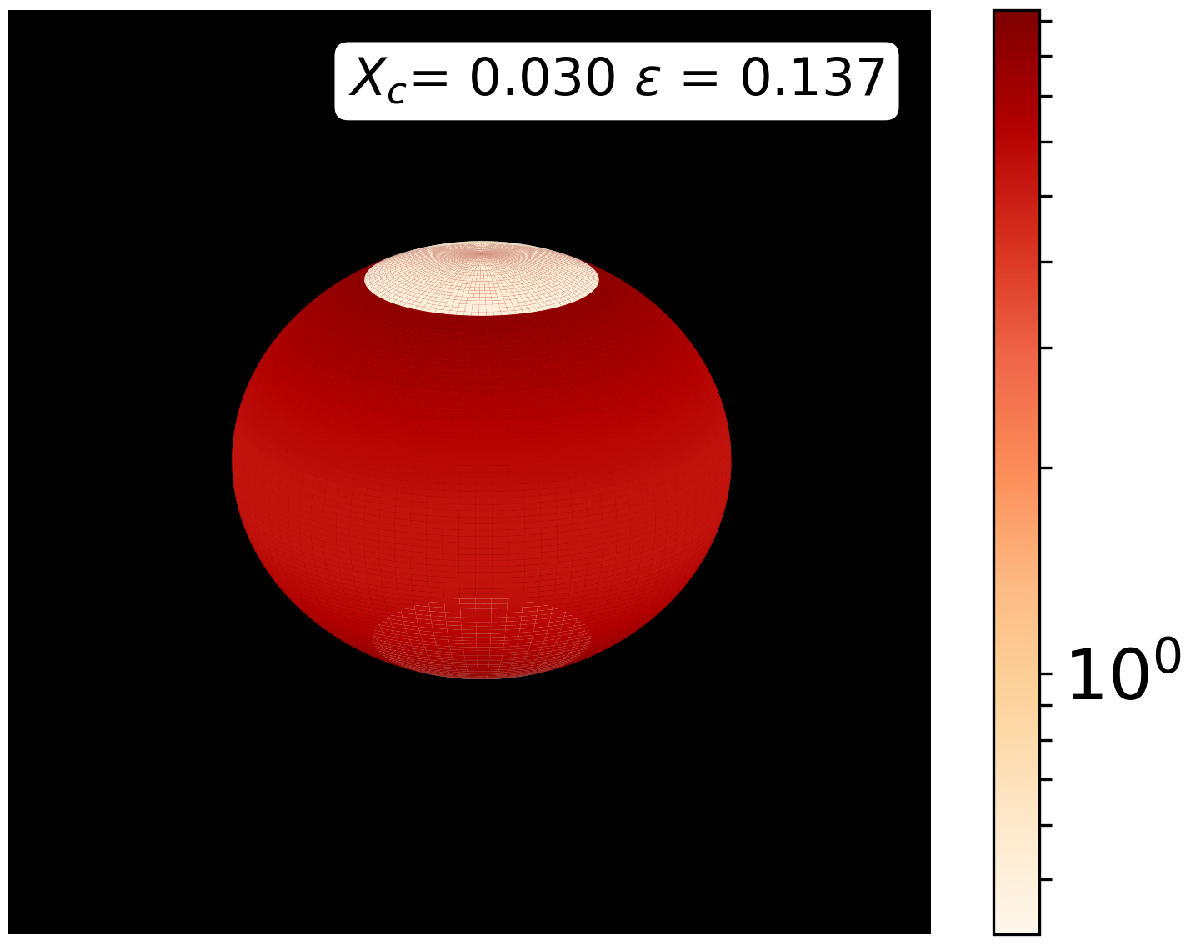}\hfill
    	\includegraphics[width=0.24\textwidth]{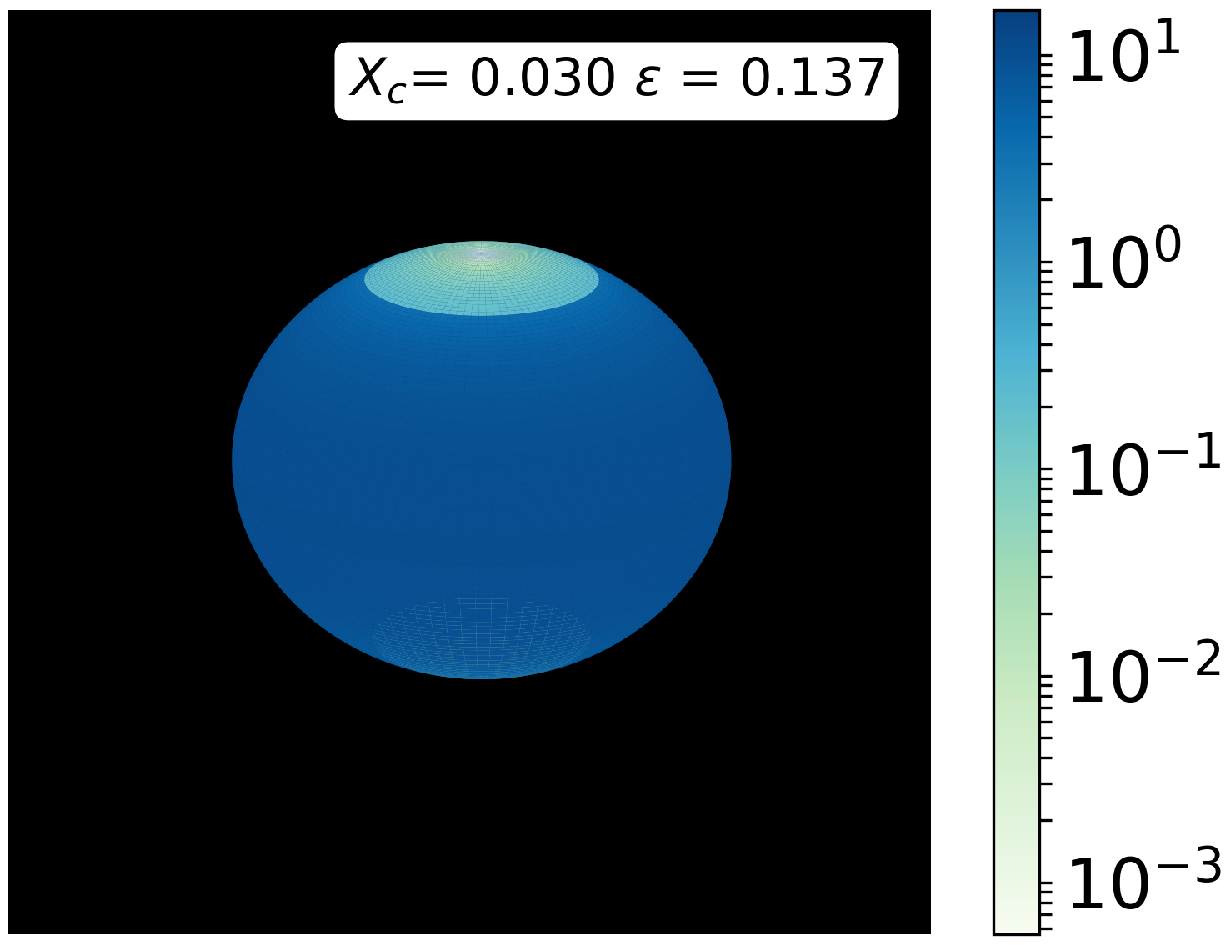}\\
	\includegraphics[width=0.24\textwidth]{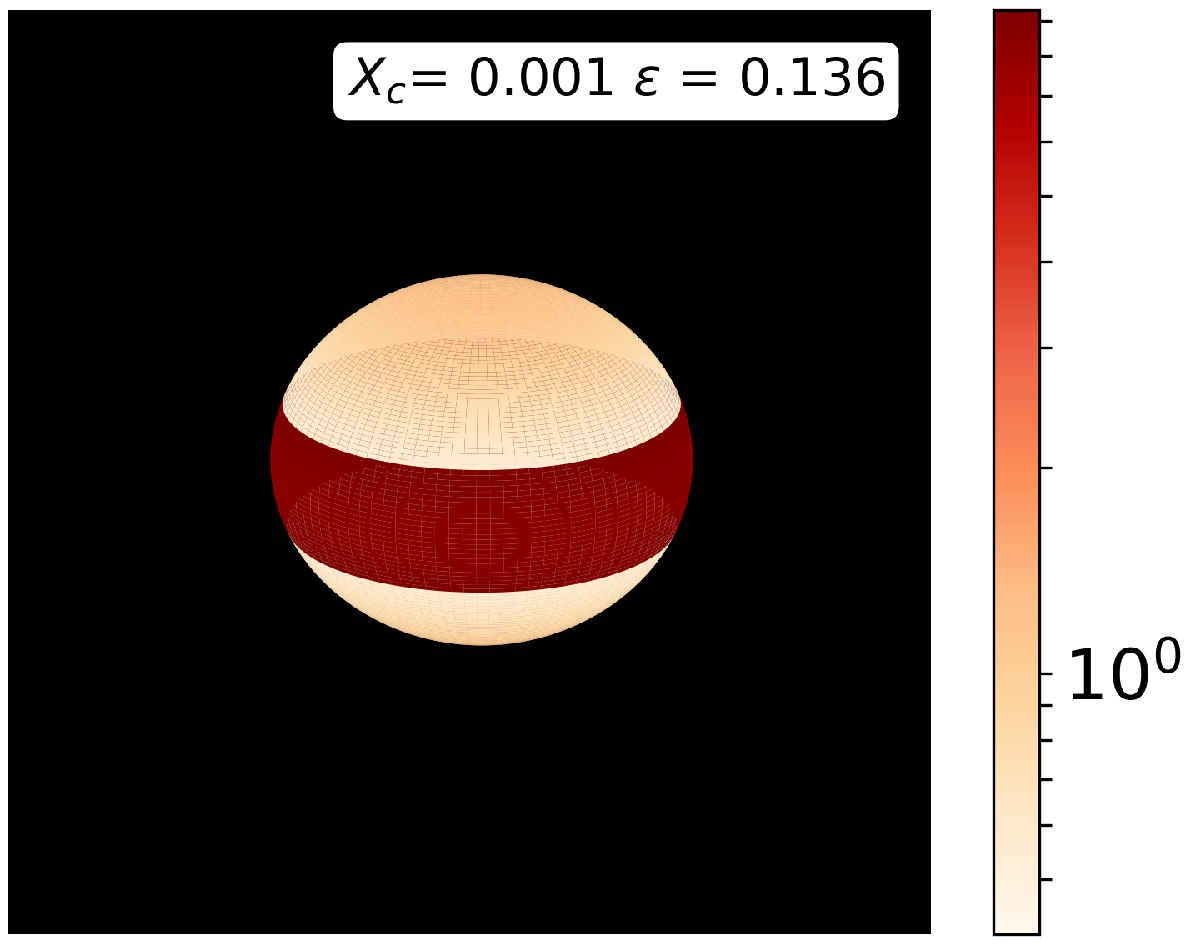}\hfill
	\includegraphics[width=0.24\textwidth]{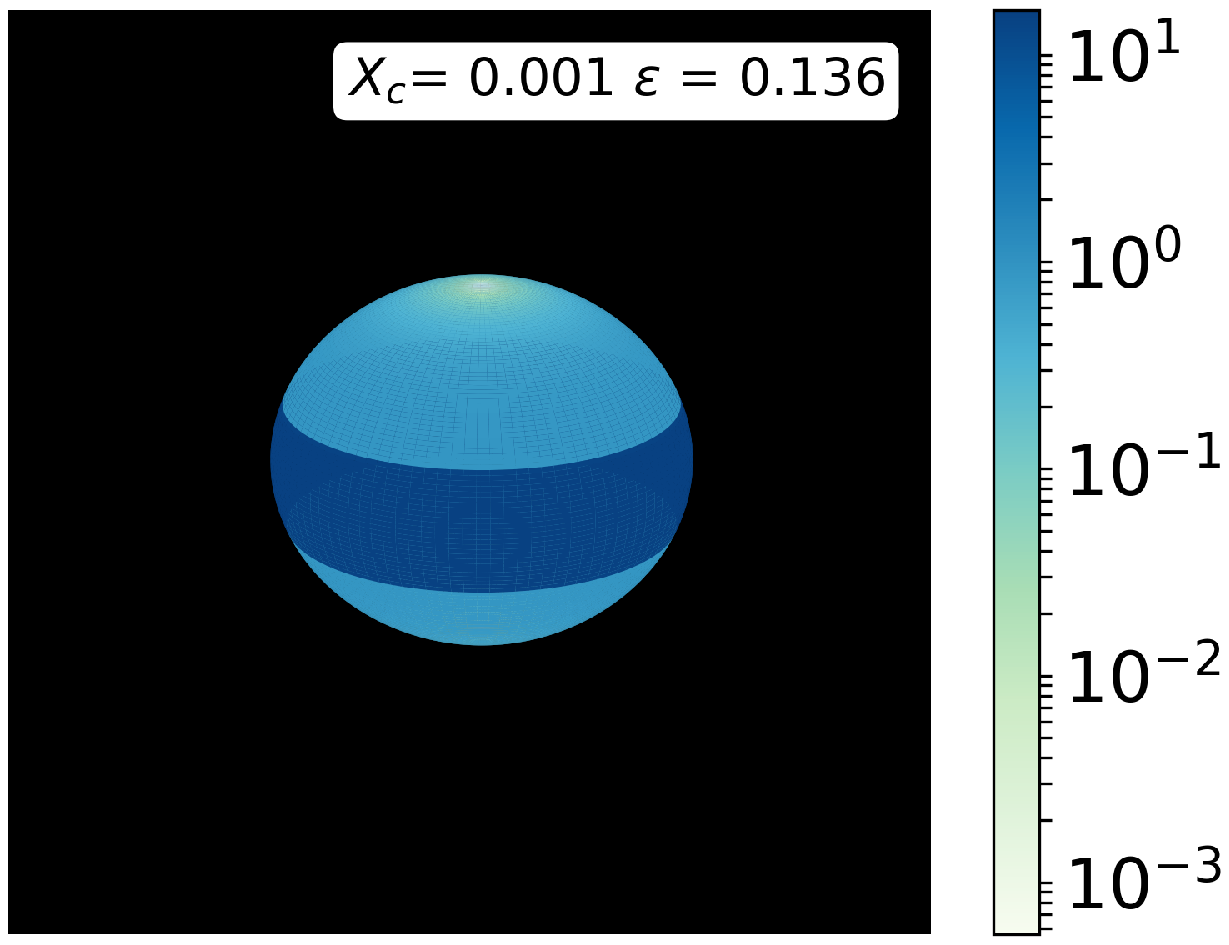}\\
  \caption{Snapshots of the local mass-flux $\dot{m}$ in units of the polar mass-flux at ZAMS $\dot{m}_{p,\rm ZAMS}$ (left) and of the local angular momentum flux $\dot{\ell}$ in units of the maximum angular momentum flux at ZAMS $\dot{\ell}_{\rm max, ZAMS}$ (right) for a 15~$M_\odot$ ESTER 2D-model with $Z=0.02$ and for $\omega_i=0.5$,  at four different time steps: $X_c\simeq 1, \ 0.3, \ 0.03 \ \rm{and} \ 0 $. The first one corresponds to ZAMS, the second corresponds to the end of Phase 1, the third is the end of Phase 2, and the last snapshot corresponds to the very end of the MS evolution. The star is viewed with an inclination $i=70\degr$.}
\label{fig:snap}
\end{figure}

Fig.~\ref{fig:mdot_Xc} and \ref{fig:Teffeq_Xc} respectively show the MS evolution of the mass-loss rate, the angular velocity ratio, and the equatorial effective temperature for an ESTER 2D-model with $M=15~M_\odot$, $Z = 0.02$, and $\omega_i=0.5$.
We find the star to start its MS evolution in a (hot-sided) SWR, that is, with an equatorial effective temperature larger than $\Tej$, until $\Xc \simeq 0.4$.  When $X_{\rm core}/X_0 \lesssim 0.4$, $T_{\rm eff, eq}$ is smaller than the effective temperature of the bi-stability jump and the star enters a TWR, then remains in this regime for the rest of the MS. In the SWR, $\log \dot{M}$ increases roughly linearly and $\omega$ increases similarly to the case with no mass loss, which is described in Sect.~\ref{constant_angmom}. The TWR can be divided into two phases of evolution. A first phase for $ 0.4 \lesssim \Xc \lesssim 0.3$ (Phase 1), where $\dot{M}$ rapidly increases while $\omega$ keeps increasing but more slowly. And a second phase for $\Xc \lesssim 0.3$, where $\log \dot{M}$ goes back to a roughly linear increase, while $\omega$ decreases. In Fig.~\ref{fig:mdot_Teff} we illustrate the MS evolution of the global mass-loss rate against the corresponding surface-averaged effective temperature $\overline{T}_{\rm eff}$ of the model. As expected (see Paper I), we find the TWR to be reached even if the surface-averaged effective temperature is larger than $T_{\rm eff}^{\rm jump}$. It underlines that the average $\Teff$ is not an appropriate quantity to determine the wind regime. 
Snapshots of the different phases along the evolution for this model are shown in Fig.~\ref{fig:snap}. We now give some explanation for the evolution of rotation and mass-loss rate in the three phases that we have identified.

\subsubsection{The single-wind regime}\label{sec:SWR}

The MS evolution of massive stars is well known to exhibit an increase of stellar luminosity $L$, and a decrease of surface averaged effective temperature $\overline{T}_{\rm eff}$. This is a consequence of the increase of central temperature and density as hydrogen burning proceeds. From (\ref{eq:m1}), (\ref{alpha}) and (\ref{k}), it is thus clear that $\dot{M}$ should increase throughout the MS, for all rotation rates, in agreement with  mass loss scaling relations \cite[e.g.][]{deJager1988,Kudritzki95}. Moreover, as shown in Paper I, the effect of rotation on $\dot{M}$ is relatively weak in the SWR \citep[see also][]{MaederMeynet2000,Petrenz2000}. The effect of the growth of $\omega$ on the evolution of $\dot{M}$ is negligible compared to the effects of secular evolution. 

Fig.~\ref{fig:mdot_Xc} (bottom) shows that, similarly to the case with no mass loss, $\omega$ increases, but more slowly. That is to be expected since angular momentum losses tend to emphasise the decrease of $\Omega_{eq}$ as evolution proceeds. Additionally, mass loss also slows down the stellar expansion as mass loss itself tends to make the star contract (see Section~\ref{sec:nonuc}). $\Omega_k$ therefore decreases more slowly when mass loss is accounted for (see Fig.~\ref{fig:Omegabk_Xc__}). 

\subsubsection{The first phase of the two-wind regime}

\begin{figure}[t]
  \centerline{\includegraphics[width=.5\textwidth]{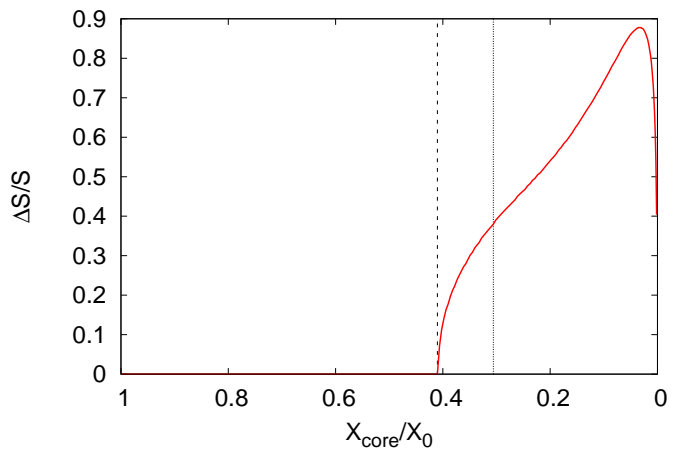}}
  \caption{Evolution of the surface fraction of the star where the effective temperature is lower than $\Tej$, for a $15M_\odot$ star with $Z = 0.02$ and for $\omega_i=0.5$. The black dashed line marks the evolutionary time at which the star reaches a TWR. The black dotted line marks the threshold between Phase 1 and 2.}
\label{fig:surface_fraction}
\end{figure}

For our 15~M$_\odot$ model, the first phase of the TWR phase corresponds to the  $0.4 \lesssim \Xc \lesssim 0.3$ time period. During this phase, the evolution of $\omega$ and $\dot{M}$ are tightly coupled. Indeed, at $\Xc \simeq 0.4$, $ T_{{\rm eff}, eq} \simeq \Tej$, and the region where the mass-flux is increased due to the local reach of bi-stability limit, is confined to the equator. At this stage, the mass and angular momentum losses are not strong enough to prevent $\omega$ from increasing. The growth is only reduced. The two phenomena controlling the evolution of $\dot{M}$ are the nuclear evolution of the star and the evolution of $\omega$. In this phase, both tend to make $\theta_{\rm jump}$ migrate poleward. Fig.~\ref{fig:surface_fraction} shows the evolution of the surface fraction of the star, $\Delta S / S$, where the effective temperature is lower than $\Tej$, that is, the portion of enhanced mass-flux due to bi-stability. In this phase, $\Delta S / S$ increases, thus leading to an increase of the mass and angular momentum loss rates. This leads to a faster decrease of $\Omega_{eq}$ and a slower decrease of the Keplerian angular velocity $\Omega_k$ (Fig.~\ref{fig:Omegabk_Xc__}) compared to models without mass loss. This process lasts until the mass and angular momentum loss rates are sufficient for $\omega$ to start decreasing. At this stage, the star enters the second phase of the TWR.

\subsubsection{The second phase of the two-wind regime}\label{sec:TWRphase2}

For this stellar model, the second phase of the TWR phase corresponds to $\Xc \lesssim 0.3$, that is, from the point where $\omega$ starts decreasing. As $\omega$ decreases we would expect $\theta_{\rm jump}$ to migrate back to the equator. However, $\Delta S/S$ and thus $\dot{M}$ still increase during this phase. Obviously, nuclear (secular) evolution effects still dominate over the consequences of the $\omega$ decrease. Hence, $\theta_{\rm jump}$ still migrates poleward. As $\omega$ decreases in this phase, the surface flattening $\epsilon$ also decreases. The effect of secular nuclear evolution becomes predominant over the effect of rotation on the evolution of equatorial opacity. This leads to a new increase of $\Gamma_{eq}$ almost until the end of the MS (see Fig.~\ref{fig:Gamma_eq}).

After the hook near end of the MS, the star contracts leading to an increase of the mean effective temperature, thus to the migration of $\theta_{\rm jump}$ towards the equator. It hence leads to a decrease of $\dot{M}$ and to an ultimate increase of $\omega$ just before the end of the MS (see Fig.~\ref{fig:mdot_Xc}).



\subsubsection{Other masses}

\begin{figure*}[t]
\includegraphics[width=.5\textwidth]{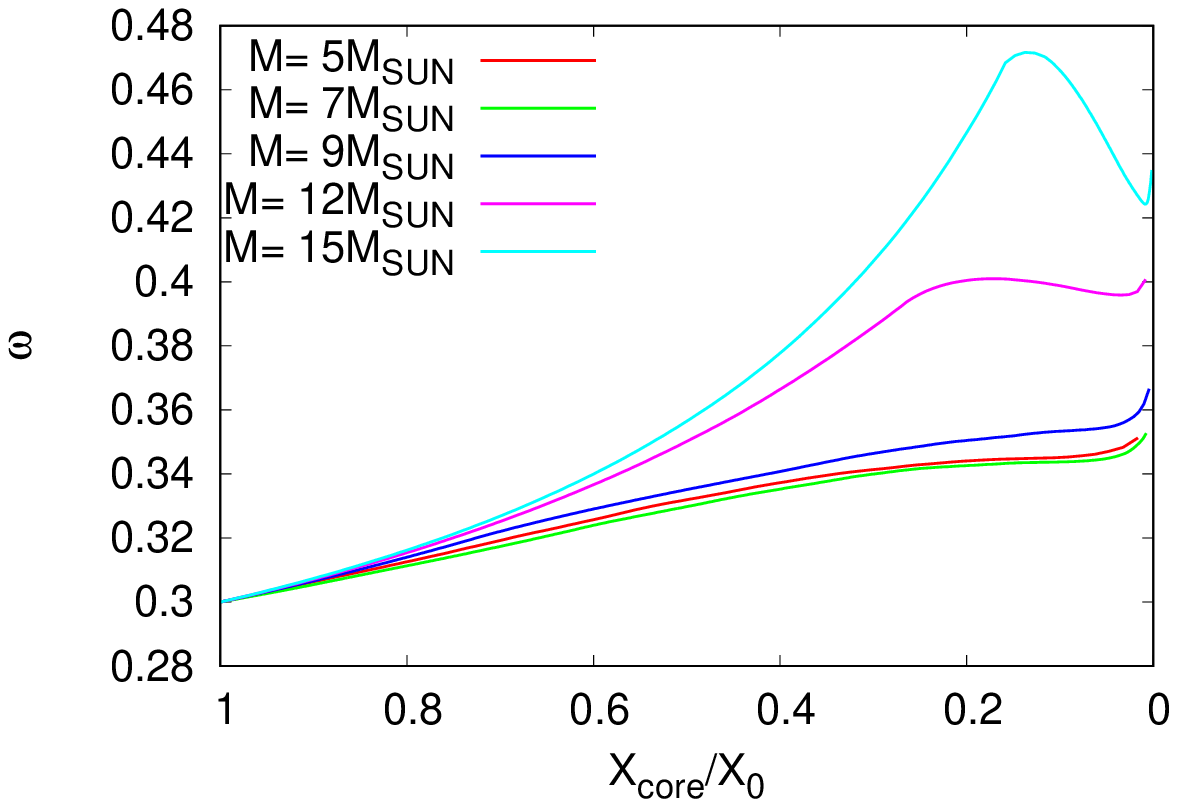}\hfill
\includegraphics[width=.5\textwidth]{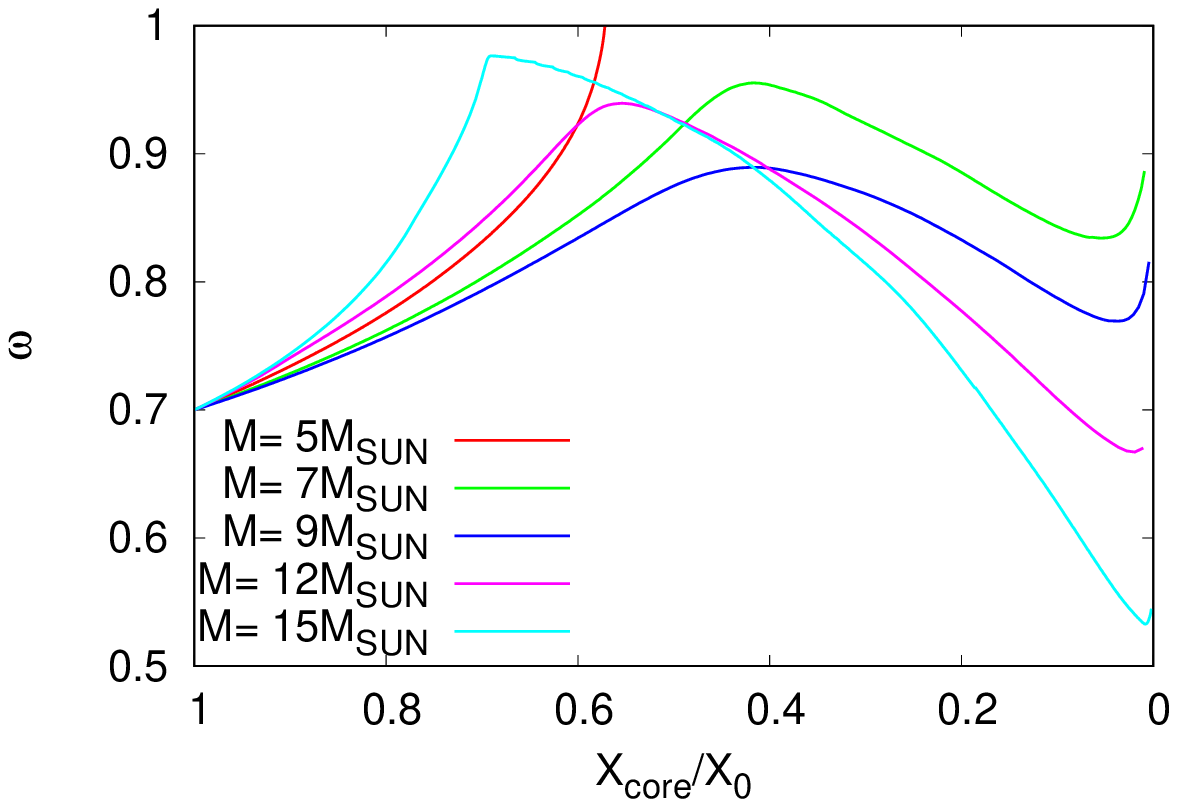}\hfill
\caption{Angular velocity at the equator in units of the equatorial critical angular velocity $\omega=\Omega_{eq}/\Omega_k$ as a function of the fractional abundance of hydrogen in the convective core $X_{\rm core}/X_0$ for different initial stellar masses and for $Z=0.02$, assuming a ZAMS angular velocity ratio $\omega_i=0.3$ (left) and $\omega_i=0.7$ (right).}
\label{fig:Omegabk_Xc_mloss_}
\end{figure*}

ESTER predictions for a given initial angular velocity ratio $\omega_{i}=0.3$ (left) and $\omega_i=0.7$ (right), and for various stellar masses, are shown in Fig.~\ref{fig:Omegabk_Xc_mloss_}. It turns out that the evolution of $\omega$ is a rather complicated function of the mass and the initial angular velocity, at least for the mass range of our models, namely $7\infapp M/M_\odot \infapp 15$. This complexity comes from the fact that the local effective temperature of such stars is often close to the effective temperature of the bi-stability jump. Hence, the stars are prone to either start their MS evolution in a hot-sided SWR (i.e. $T_{\rm eff} > T_{\rm eff}^{\rm jump}$ for all $\theta$) and possibly reach a TWR followed by a cold-sided SWR (i.e. $T_{\rm eff} < T_{\rm eff}^{\rm jump}$ for all $\theta$), or start in a TWR and reach a cold-sided SWR. Hence, the evolution of $\omega$ may drastically vary when initial conditions are changed.

The modelling of more massive stars is certainly more predictable. Indeed, such stars are always in a hot-sided SWR at ZAMS, and above a certain mass, the expansion and evolution of rotation is not sufficient to reach a TWR. Additionally, more massive stars lose a lot more mass and angular momentum because of their stronger radiation-driven wind, and thus may exhibit a decreasing $\omega$ all along the MS. Unfortunately, for numerical reasons, the ESTER code presently does not allow us to study the nuclear evolution of models with masses much higher than $\sim 20~M_\odot$. 


\section{Conclusions}\label{sec:conclu}

This work presents a study of the rotational evolution of rapidly rotating early-type stars with the 2D code ESTER that we have updated to follow the decrease of core hydrogen content as a proxy for time evolution along the MS. 
We first investigated the rotational evolution of intermediate-mass stars with masses in-between $5$ and $7~M_\odot$. The evolution of such stars along the MS is relatively simple because at first order it can be modelled with constant angular momentum, that is, without considering any mass loss. We have shown that, because of stellar expansion as well as the redistribution of mass in the stellar interior, critical rotation can be reached before the end of the MS provided the initial angular velocity ratio $\omega_i= \Omega_{eq,i}/\Omega_k$ is high enough. The minimum value of $\omega_i$ required to reach critical rotation is smaller for the more massive stars. In this mass range, no star reaches the critical angular velocity during the MS if its initial angular velocity is less than $50\%$ critical. 

At masses $M \geq 7~M_\odot$, our models show that the rotational evolution is more complex. Indeed, such stars are subject to radiation-driven winds, inducing mass and angular momentum losses, which substantially impact the evolution of the angular velocity ratio $\omega$. Unfortunately, the magnitude of mass-loss rates obtained with CAK-based prescriptions are rather uncertain, because of loosely constrained parameters and the neglect of wind inhomogeneities (see Paper I for more details). More accurate mass-loss rates could be derived from recent hydrodynamically consistent non-local thermodynamical equilibrium stellar atmosphere models \citep{Sander2017}, but such models will certainly remain 1D for some time.  Nevertheless, we used the mass-flux prescription detailed in Paper I, based on the modified CAK theory \citep{PPK}, where we include, \textit{locally,} the bi-stability limit. For now, the existence of this jump is still challenged by the observations \citep[e.g.][]{Markova2008}, but their conclusions are not clear cut. From the point of view of the models, the bi-stability jump has important consequences and hence cannot be overlooked. Indeed, in 2D-models, the bi-stability limit imposes the existence of two wind regimes that have rather different effects on the rotational evolution of massive stars. Stars are in a single-wind regime (SWR) when, at all colatitudes, their effective temperature is either larger or smaller than the effective temperature of the bi-stability jump. This wind regime exhibits a maximum mass-flux at the poles. However, when there is a colatitude $\theta_{\rm jump}$ where the effective temperature equals the effective temperature of the bi-stability jump, the star is in a two-wind regime (TWR). In this regime, the mass-flux is maximum in the equatorial region and holds a discontinuity at $\theta_{\rm jump}$.

The main result of this study is that the rotational evolution of massive stars with masses in-between 7 and $15~M_\odot$ is very dependent on their initial conditions at ZAMS, namely their mass and initial angular velocity. Indeed, their effective temperature is rather close to the effective temperature of the bi-stability jump.  The initial angular velocity therefore controls whether stellar models in this mass range start their MS evolution in a SWR or in a TWR. 
The different wind regimes and transitions between these regimes as evolution proceeds, can have very different effects on the rotational evolution of such stars, hence the importance of the ZAMS initial conditions.
The most common transition from one regime to another is the transition from a hot-sided SWR to a TWR. We have detailed the rotational evolution of a 2D-ESTER model with M=15~M$_\odot$ initially rotating with an angular velocity ratio $\omega_i=0.5$. The SWR-TWR transition is followed by two phases of evolution of the angular velocity ratio. These two phases correspond to a rapid (and continuous) increase of the mass loss and angular-momentum-loss rates leading to a slower growth of $\omega$ (compared to the SWR), followed by the decrease of $\omega$ accompanied by a slow increase of the mass and angular momentum loss rates. 

Another important result is that, using 2D models, and despite the use of \cite{Vink2001} to calibrate our mass loss prescription in the non-rotating case, we do not find a discontinuity of the global mass-loss rate at the bi-stability limit as in 1D models. Instead, the discontinuity lies on the local mass-flux at the equator, which then migrates poleward as evolution proceeds leading to a continuous increase of the global mass-loss rate in the TWR. Because of the local nature of the bi-stability limit in 2D models, the latter could be reached at various rotation rates and stellar masses, thus various surface-averaged effective temperatures. In this work, we have chosen the effective temperature of the bi-stability limit to be that of \cite{Vink2001}, namely roughly $25 {\rm kK}$. Taking a lower value, for instance $ \sim 20{\rm kK}$ from \cite{Petrov2016} would shift the stellar models that may reach a TWR during the MS towards less massive stars. Models that remain in the requisite mass range would reach a TWR later on the MS. The effects of a reduced effective temperature of the bi-stability limit might become particularly important for later
evolutionary phases of more massive stars (B-supergiants).

All in all, we find that radiation-driven mass and angular momentum losses are responsible for either a slower growth or a decrease of the angular velocity ratio $\omega$ along the MS. This often prevents massive stars from reaching critical rotation before the end of the MS.

The present work calls for new investigations to confirm the present results, in particular in the modelling of the mass losses. This is an old problem, which we here face through the modelling of the force multipliers parameters (see references in Paper I). It is known that these parameters depend on the metallicity of the atmosphere sourcing the wind, but the changes to be applied for modelling a population III star, for instance, are still quite uncertain \citep[e.g.][]{Meynet2008}. The attempt of \cite{Georgy2013} finds that a change in metallicity does not bring any significant change in the rotational evolution of a $15~M_\odot$ model, but their models are one-dimensional. We also note that there is strong evidence that massive stars undergo periods of super-Eddington conditions \citep{Quataert2016} in which they could exhibit continuum-driven winds, sufficiently strong to sustain significant mass loss, even for non-metallic stars \citep[e.g.,][]{Owocki2004, Smith2006, Owocki2017}.

Beside these open questions related to the influence of metallicity, the present work also calls for an investigation of the coupling between mass losses and interior flows. Indeed, the rotational evolution of the star implies the presence of meridional flows that carry chemicals from the core to the surface. Presently, the 2D evolution given by ESTER models is a succession of steady states, with steady flows, that are hardly usable to predict the transport of chemicals inside the star. New studies, that will include viscous stresses everywhere in the star, and not only in Ekman layers as presently, are needed to give a full 2D-view of the transport process in a radiative envelope and shed new light on the long standing question of rotational mixing.

\begin{acknowledgements}
We are particularly grateful to Joachim Puls for his detailed reading, comments and suggestions on the original manuscript. We thank Georges Meynet and Fabrice Martins for enlightening discussions. We are grateful to Sylvia Ekström for providing information to validate our scheme for main sequence temporal evolution. We thank CALMIP -- the computing centre of Toulouse University (Grant 2017-P0107). M. Rieutord  acknowledges the strong support of the French Agence Nationale de la Recherche (ANR), under grant ESRR (ANR-16-CE31-0007-01), and of the International Space Science Institute (ISSI) for its support to the project ``Towards a new generation of massive star models'' lead by Cyril Georgy. F. Espinosa Lara acknowledges the financial support of the
Spanish MINECO under project ESP2017-88436-R.
\end{acknowledgements}

\bibliographystyle{aa}
\bibliography{bibnew}

\begin{appendix}\label{AppendixA}
\section{Rate of change of the mass fraction of hydrogen in the stellar convective core}

In this appendix we calculate the rate of change of the mass fraction of hydrogen in the core. We write the local conservation of hydrogen (which is equivalent to Eq.~\ref{Xdot}) as 

\begin{equation}\label{Xconserv}
\frac{\partial }{\partial t}(\rho X) + {\rm div}(\rho X \bv) = -\frac{4m_{\rm p}}{Q}\rho \epsilon_{*} \ .
\end{equation}
In order to get an expression for $dX_{\rm core}/dt$, we calculate the rate of change of the mass of hydrogen in the core, namely

\begin{equation}
\frac{d}{dt}(M_{\rm core} X_{\rm core}) = \frac{d}{dt} \int_{\rm core} \rho X dV \ .
\end{equation}
This equation can be re-written using Leibniz's rule as

\begin{equation}
\frac{d}{dt}(M_{\rm core} X_{\rm core}) =  \int_{\rm core} \frac{\partial}{\partial t}(\rho X) dV + \int_{\partial \rm core} \rho X \bv_c \cdot \boldsymbol{dS} \ ,
\end{equation}
where $\bv_c$ is the speed of the core boundary. Using equation~(\ref{Xconserv}) and the fact that we consider the core to be fully mixed, that is, $X$ is homogeneous in the core,

\begin{equation}\label{prevLeibniz}
\begin{aligned}
\frac{d}{dt}(M_{\rm core} X_{\rm core}) &=   -\frac{4m_{\rm p}}{Q}L_{\rm core} \\ & + X_{\rm core} \left( - \int_{\rm core} {\rm div}(\rho \bv) dV +  \int_{\partial \rm core} \rho  \bv_c \cdot \boldsymbol{dS} \right) \ .
\end{aligned}
\end{equation}
Using Leibniz's rule again, equation~(\ref{prevLeibniz}) becomes

\begin{equation}
\begin{aligned}
\frac{d}{dt}(M_{\rm core} X_{\rm core})& =   -\frac{4m_{\rm p}}{Q}L_{\rm core} + X_{\rm core}\frac{d}{dt} \int_{\rm core} \rho dV \\
&= -\frac{4m_{\rm p}}{Q}L_{\rm core} + X_{\rm core}\frac{dM_{\rm core}}{dt} \ ,
\end{aligned}
\end{equation}
which finally leads to

\begin{equation}
\frac{dX_{\rm core}}{dt}=-\frac{4 m_{\rm p}}{Q} \frac{L_{\rm core}}{M_{\rm core}} \ .
\end{equation}

\end{appendix}

\end{document}